\documentclass[tightenlines,superscriptaddress,twocolumn,preprintnumbers,amssymb,nofootinbib]{revtex4-1}
\usepackage{amsmath, amsthm, amssymb, mathrsfs, bm} 
\usepackage{graphics,bm}
\usepackage{epsfig}
\usepackage{graphicx}
\usepackage{amsmath}
\usepackage{wrapfig}
\usepackage{color}
\usepackage{tikz}
\RequirePackage{mathtools}
\RequirePackage{subfigure}
\usetikzlibrary{arrows,shapes}
\usetikzlibrary{trees}
\usetikzlibrary{matrix,arrows} 			
\usepackage[bottom]{footmisc}											
\usetikzlibrary{positioning}				
\usetikzlibrary{calc,through}				
\usepackage{pgffor}							

\usetikzlibrary{decorations.pathmorphing}	
\usetikzlibrary{decorations.markings}
\tikzset{
    sigmaCT/.style={draw=black, postaction={decorate},
        decoration={markings,mark=at position .99 with {\arrow[draw=black]{>}},mark=at 		 position .99 with {\arrow[
draw=black]{<}}}},
    pionCT/.style={dashed,draw=black, postaction={decorate},
        decoration={markings,mark=at position .99 with {\arrow[draw=black]{>}},mark=at position .99 with {\arrow[draw=black]{<}}}},    
    fermionCT/.style={draw=black, postaction={decorate},
        decoration={markings,mark=at position .5 with {\arrow[draw=black]{>}},mark=at position .99 with {\arrow[draw=black]{>}},mark=at position .99 with {\arrow[draw=black]{<}}}},    
    fermion/.style={draw=black, postaction={decorate},
        decoration={markings,mark=at position .55 with {\arrow[draw=black]{>}}}},
    fermionbar/.style={draw=black, postaction={decorate},
        decoration={markings,mark=at position .55 with {\arrow[draw=black]{<}}}},
    pion/.style={dashed,draw=black, postaction={decorate}},
    sigma/.style={draw=black, postaction={decorate}}
}

\newcommand{\beq}{\begin{equation}}
\newcommand{\eeq}{\end{equation}}
\newcommand{\bqa}{\begin{eqnarray}}
\newcommand{\eqa}{\end{eqnarray}}

\newcommand{\os}{\text{\tiny OS}}
\newcommand{\ms}{\overline{\text{\tiny MS}}}

\begin{document}

\title{On-shell versus curvature mass parameter fixing schemes in the quark-meson model and its phase diagrams}
\author{Suraj Kumar Rai}
\email{surajrai050@gmail.com}
\affiliation{Department of Physics, University of Allahabad, Prayagraj, India-211002}
\affiliation{Department of Physics, Acharya Narendra Deo Kisan P.G. College, Babhnan Gonda, India-271313}
\author{Vivek Kumar Tiwari}
\email{vivekkrt@gmail.com}
\affiliation{Department of Physics, University of Allahabad, Prayagraj, India-211002}
\date{\today}

\begin{abstract}

We compute and compare the effective potential and phase structure for the quark-meson model in an extended mean-field
approximation (e-MFA) when vacuum one loop quark fluctuations are included and the model parameters are fixed 
using different renormalization prescriptions.When the quark one loop vacuum divergence is regularized under the minimal subtraction scheme,the model setting of the parameter fixing using the curvature masses of the scalar and pseudo-scalar mesons, has been termed as the quark-meson model with the vacuum term (QMVT).However,this prescription becomes inconsistent when we notice that the curvature mass is akin to defining the meson mass by the self-energy evaluation at vanishing momentum.In this work,we apply the recently reported exact prescription of the on-shell parameter fixing,
to that version of quark-meson (QM) model where the two quark flavors are coupled to the  
eight mesons of the $ SU(2)_{L} \times SU(2)_{R} $ linear sigma model with iso-singlet $ \sigma $ ($\eta$), 
iso-triplet $ \vec{a_{0}} $ ($ \vec{\pi} $) scalar(pseudo-scalar) mesons.The model then becomes,
the renormalized quark-meson (RQM) model where physical (pole) masses of mesons and pion 
decay constant, are put into the relation of the running mass parameter and couplings by using 
the on-shell and the minimal subtraction renormalization schemes.The vacuum effective potential plots,the phase diagrams and the order parameter temperature variations for both the RQM model and QMVT model,are exactly identical for the $m_\sigma=$616 MeV.The vacuum effective potential when the $m_\sigma < $ 616 MeV,is deepest for the QMVT model and 
an interesting trend reversal is observed when the effective potential of the RQM model becomes deepest  for the $m_\sigma > $ 616 MeV.We find similar $ m_{\sigma} $ dependent differences in the nature of the RQM and QMVT model phase diagrams and the  order parameter temperature variations.Furthermore, $ SU(2)_{A}$ chiral and $ U(1)_{A}$ axial symmetry breaking/restoration and their interplay can also be investigated in this framework.

\end{abstract}
\keywords{ Dense QCD,
chiral transition,}

\maketitle
\section{Introduction}
  
Strong interaction theory motivated, first quantum chromodynamics (QCD) schematic phase diagram, appeared in 1970s \cite{Cabibbo75} which projected a low temperature (low baryonic density) confined phase of hadrons and a high temperature (zero baryonic density) or high baryonic density (zero temperature) phase of quarks and gluons \cite{SveLer,Mull,Ortms,Riske}.Mapping out the QCD phase diagram in all its details is still a very active area of current research as it is not very well understood.The first-principle lattice QCD simulations\cite{AliKhan:2001ek,Digal:01,Karsch:02,Fodor:03,Allton:05,Karsch:05,Aoki:06,Cheng:06,Cheng:08}provide us 
valuable information for the QCD phase transition but the real progress in the lattice QCD calculations,gets 
seriously bogged down by the QCD action becoming complex due to the fermion sign problem\cite{Karsch:02}when baryon density/chemical potential becomes nonzero.The phenomenological models developed with the effective degrees of 
freedom\cite{Alf,Fukhat},are of great help in mapping out the phase diagram in the regions inaccessible to the lattice simulations. 

The QCD Lagrangian for the two flavor of massless quarks has the $ SU(2)_{L+R} \times SU(2)_{L-R} $ symmetry.The axial(A=L-R) part of the symmetry called the chiral symmetry,gets spontaneously broken in the low energy hadronic vacuum of the QCD as the  chiral condensate forms and one gets three massless pions as Goldstone bosons.The chiral symmetry gets explicitly broken as well,due to the small mass of the u and d quarks and we find light pions in the nature.The $ SU(2)_{L} \times SU(2)_{R} $ linear sigma model provide us a good framework \cite{Roder,fuku11,grahl} in which the  chiral $ SU(2)_{A}$ symmetry and the axial $U(1)_{A}$ symmetry breaking and restoration both can be investigated in great detail as it enables the construction of chiral invariant combinations using the chiral partners from the iso-singlet $ \sigma $ and the iso-triplet $ \vec{a_{0}} $ scalar mesons to the iso-singlet $\eta$ and iso-triplet $ \vec{\pi} $ pseudo-scalar mesons.Coupling these eight scalar and pseudo-scalar meson degrees of freedom to the two flavor of quarks, we get the QCD-like framework of the quark-meson (QM) model in which we can compute and explore the QCD phase diagram.

Furthermore,at low temperatures and densities,the confinement of quarks inside the hadrons also,gets implemented by the introduction of the Polyakov loop where the QCD confinement is mimicked in a statistical sense by coupling the chiral models to a  constant background $SU(N_{c})$ gauge field $A_{\mu}^{a}$ \cite{Polyakov:78plb,fuku,benji, Pisarski:00prd,Vkt:06}.Using phenomenological Polyakov loop potential \cite{ratti,fuku2},the free energy density from the gluons is added to the QM model and it becomes the PQM model.Several investigations of the QCD phase structure/phase diagram, have already been done in the chiral models \cite{scav, Rischke:00,jakobi,Herpay:05,Herpay:06,Herpay:07,Kovacs:2006ym,kahara,Bowman:2008kc,Fejos,Jakovac:2010uy,koch,marko}, two and three flavor QM model \cite{mocsy,bj,Schaefer:2006ds,Schaefer:09} and PQM model \cite{SchaPQM2F,SchaPQM3F,Mao,TiPQM3F}.  

Under the “no-Dirac sea” or standard mean field approximation (s-MFA),fermionic vacuum fluctuations and renormalization issues are neglected altogether\cite{scav,mocsy,bj,Schaefer:2006ds,SchaPQM2F,kahara,Schaefer:09,SchaPQM3F,Mao,TiPQM3F} assuming that the redefined meson potential parameters would reabsorb their effects.This familiar setting of the QM model gives inconsistent result as in the chiral limit,one gets a first-order chiral phase transition at zero baryon densities which is at odds with the general theoretical arguments\cite{rob,hjss}.The above inconsistency is remedied with the proper treatment of the Dirac sea as first proposed in the Ref. \cite{vac} and later several research papers \cite{lars,guptiw,schafwag12,chatmoh1,TranAnd,vkkr12,chatmoh2,vkkt13,Herbst,Weyrich,kovacs,zacchi1,zacchi2,Rai} worked out the detailed impact of including the  quark one loop vacuum correction in the two and three flavor QM/PQM model.In these publications,it has been the standard procedure to identify the pion decay constant with the vacuum expectation value of the sigma mean field and to use the $\sigma$ and $\pi$ meson curvature mass,defined by the second derivative of the thermodynamic potential at its minimum,for fixing the model parameters where the minimal subtraction scheme has been used to properly regularize the quark one loop vacuum divergence.Since the effective potential generates the n-point functions of the theory at vanishing external momenta,the curvature mass is akin to defining the meson mass by the evaluation of self-energy at zero momentum \cite{laine,Adhiand1,BubaCar,Naylor,fix1}.This consideration renders the above parameter fixing procedure inconsistent.In order to make comparisons and quantify the effect of the parameter fixing with the curvature meson masses,we have named this model setting as the quark-meson model with vacuum term (QMVT).

The six parameters, $\lambda_{1}$, $\lambda_{2}$, t'Hooft coupling $c$, mass parameter $m^{2}$, explicit symmetry breaking strength $h$ and Yukawa coupling $g$ of the QM model Lagrangian are determined by the physical values of $m_{\sigma}$,$m_{\pi}$,eta meson mass $m_{\eta}$,isotriplet scalar meson mass $m_{\vec{a_{0}}}$,constituent quark mass $m_{q}$ and pion decay constant $f_{\pi}$.It is to be noted that,in most renormalization procedures,radiative corrections to the physical quantities change their tree level relations to the parameters of the Lagrangian.Thus the use of tree level values of the parameters, in the effective potential calculation, becomes inconsistent.The 
$\overline{\text{MS}}$ scheme running parameters 
depend on the renormalization scale $\Lambda$,whereas the on-shell parameters have their tree-level values. 
The correct renormalization prescription allows us to calculate the counterterms both in the $\overline{\text{MS}}$
scheme and in the on-shell scheme and then the renormalized parameters of the two schemes get connected.The effective potential is then calculated using the modified minimal subtraction procedure where the relations between the running parameters and the on-shell parameters (physical quantities) are used as input \cite{Adhiand1}.Adhikari and collaborators in a series of papers \cite{Adhiand1,Adhiand2,Adhiand3,asmuAnd} used this renormalization prescription to correctly account for the effect of Dirac sea in the context of QM model where $O(4)$ sigma model has been used for 
the mesonic degree of freedom (iso-singlet scalar meson $\sigma$ and iso-triplet $\pi$).In the present work, we are applying this prescription of the on-shell parameter fixing to that version of the quark-meson (QM) model in which the two flavor of quarks are coupled to the  
eight mesons of the $ SU(2)_{L} \times SU(2)_{R} $ linear sigma model with iso-singlet $ \sigma $, 
iso-triplet $ \vec{a_{0}} $ scalar mesons and iso-singlet $\eta$, iso-triplet $ \vec{\pi} $ pseudo-scalar mesons.We have termed this model setting as the renormalized quark-meson (RQM) model which has the advantage of providing us the framework in which apart from the $ SU(2)_{A}$ chiral, we can investigate the $ U(1)_{A}$ axial symmetry breaking and restoration also together with the interplay of axial $ U(1)_{A}$ and $ SU(2)_{A}$ chiral symmetry.

    The paper is arranged as follows.The brief formulation of the $ SU(2)_{L} \times SU(2)_{R} $ QM model is presented in the section II.The section III presents the calculation of the effective potential of the quark-meson model with vacuum term (QMVT) together with its parameter fixing procedure which uses the curvature masses of the scalar and pseudo-scalar mesons.  
    The on-shell scheme counterterms and self-energy calculations are presented in the section IV-A.The 
    relations between the physical quantities and the running parameters, are derived in the section IV-B,the derivation of the effective potential in the RQM model is also presented in the section IV-C.The result and discussion is presented in the section V.Finally summary
    and conclusion is presented in the section VI.

\section{Model Formulation}
The $ SU(2)_{L} \times SU(2)_{R} $ quark-meson model formulation will be presented
in this section.
In the two flavor quark meson chiral linear sigma model,two light quarks and  $SU_V(2)\times SU_A(2)$ symmetric meson fields are coupled together.The Lagrangian of the model \cite{Roder,fuku11,grahl} is written as
\bqa
\nonumber
{\cal L_{QM}}&=&\bar{\psi}[i\gamma^\mu \partial_\mu-g t_0(\sigma+i\gamma_5 \eta)\\
&&-g \vec t\cdot(\vec a+i\gamma_5 \vec\pi)]\psi+\cal{L(M)},
\label{lag}
\eqa
where $\psi$ is a color $N_c$-plet, a four-component Dirac spinor as well as a flavor doublet 
\bqa
\psi&=&
\left(
\begin{array}{c}
u\\
d
\end{array}\right)\;.
\eqa
The Lagrangian for meson fields is \cite{Roder} 
\bqa
\nonumber
\label{lag11}
\cal{L(M)}&=&\text{Tr} (\partial_\mu {\cal{M}}^{\dagger}\partial^\mu {\cal{M}}-m^{2}({\cal{M}}^{\dagger}{\cal{M}}))\\
\nonumber
&&-\lambda_1\left[\text{Tr}({\cal{M}}^{\dagger}{\cal{M}})\right]^2-\lambda_2\text{Tr}({\cal{M}}^{\dagger}{\cal{M}})^2\\
&&+c[\text{det}{\cal{M}}+\text{det}{\cal{M}}^\dagger]+\text{Tr}\left[H({\cal{M}}+{\cal{M}}^\dagger)\right]
\label{lag1}
\eqa
here field ${\cal{M}}$ is a complex $2\times2$ matrix
\bqa
\nonumber
{\cal{M}}&=&t_a\xi_a=t_a(\sigma_a+i\pi_a)\;
\eqa
$a=$0,1,2 and 3. $t_a$ represents the 4 generators of the $U(2)$ algebra.
\bqa
\nonumber
{\cal{M}}&=&t_0(\sigma_0+i\pi_0)+\vec{t}\cdot(\vec{\sigma}+i\vec{\pi})\;\\
&=&t_0(\sigma+i\eta)+\vec t\cdot(\vec a+i\vec\pi)
\eqa
\ \ \ \ \ \ \ \ \ \ \ \ \ \ \ \ \ \  with $t_0=\dfrac{1}{2}\left(
\begin{array}{c c}
1 & 0 \\
0 & 1
\end{array}\right),$ $t_1=\dfrac{1}{2}\left(
\begin{array}{c c}
0 & 1 \\
1 & 0
\end{array}\right)$ \\ 
,\ \ \ \ \ \ \ \ \ \  \ \ \ \ \ \ \ \ \ \ \ \  \  \ $ t_2=\dfrac{1}{2}\left(
\begin{array}{c c}
0 & -i \\
i & 0
\end{array}\right),$ $t_3=\dfrac{1}{2}\left(
\begin{array}{c c}
1 & 0 \\
0 & -1
\end{array}\right).$ 
One can rewrite the Lagrangian (\ref{lag11})  in the form \cite{fuku11}
\bqa
\nonumber
{\cal{L(M)}}&=&\mbox{$1\over2$}(\partial_\mu\sigma\partial_\mu\sigma+\partial_\mu\vec{\pi}\cdot\partial_\mu\vec{\pi}+\partial_\mu\eta\partial_\mu\eta \\
&&+\partial_\mu\vec{a_0}\cdot\partial_\mu\vec{a_0})-U
\eqa
further 
\bqa
\nonumber
U&=&\dfrac{m^2}{2}(\sigma^2+{\vec\pi}^2+\eta^2+{\vec a}^2)-\dfrac{c}{2}(\sigma^2-\eta^2+{\vec\pi}^2-{\vec a}^2)\\
\nonumber
&&+\dfrac{1}{4}\left(\lambda_1+\dfrac{1}{2}\lambda_2\right)(\sigma^2+{\vec\pi}^2+\eta^2+{\vec a}^2)^2\\
\nonumber
&&+\dfrac{\lambda_2}{2}\left((\sigma^2+{\vec\pi}^2)(\eta^2+{\vec a}^2)-(\sigma\eta-{\vec\pi}\cdot{\vec a})^2\right)\\
&&-h\sigma
\eqa
The $2\times2$ matrix $H$ explicitly breaks the chiral symmetry and is chosen as 
\bqa
H=t_a h_a
\eqa
where $h_a$ are external fields.
The field $\sigma$ acquires nonzero vacuum expectation value(VEV),$\overline \sigma$, due to the spontaneous breaking of the chiral symmetry, while the other scalar and pseudo-scalar fields ($\vec a_0 ,\vec \pi,\eta$) assume zero VEV.Here the two parameters $h_0$ and $h_3$ may give rise to the explicit breaking of chiral symmetry.We are neglecting the isospin symmetry breaking,hence we choose $h_0\neq0$ and $h_3 = 0$.

The field $\sigma$ has to be shifted to $\sigma \xrightarrow \  \overline \sigma + \sigma  $ as it acquires nonzero VEV. At the tree level,the expression of meson masses are \cite{Roder}
\bqa
\label{m1}
m^2_\sigma&=&m^2-c+3(\lambda_1+\frac{\lambda_2}{2})\overline  \sigma^2\;,\\
m^2_{a_0}&=&m^2+c+(\lambda_1+\frac{3\lambda_2}{2})\overline \sigma^2\;,\\
m^2_\eta&=&m^2+c+(\lambda_1+\frac{\lambda_2}{2})\overline \sigma^2\;,\\
m^2_\pi&=&m^2-c+(\lambda_1+\frac{\lambda_2}{2})\overline \sigma^2\;,\\
m_q&=& \frac{g \ \overline \sigma}{2}
\label{m4}
\eqa
Using (\ref{m1})--(\ref{m4}),the parameters of the Lagrangian (\ref{lag1}) are obtained as  
\bqa
\label{para1}
\lambda_1&=&\dfrac{m^2_\sigma+m^2_\eta-m^2_{a_0}-m^2_\pi}{2\overline{\sigma}^2}\;,\\
\lambda_2&=&\dfrac{m^2_{a_0}-m^2_{\eta}}{\overline{\sigma}^2}\;,\\
m^2&=&m^2_\pi+\dfrac{m^2_\eta-m^2_\sigma}{2}\;,\\
c&=&\dfrac{m^2_\eta-m^2_\pi}{2}\;,\\
\frac{g^2}{4}&=&\frac{m_{q}^2}{ \overline \sigma^2}
\label{para4}
\eqa
and the tree level effective potential is written as,
\bqa
\label{effpot}
U(\overline \sigma)=\frac{1}{2}m^2\overline \sigma^2-\frac{1}{2}c\overline \sigma^2+\frac{1}{4}\left(\lambda_1+\frac{1}{2}\lambda_2\right)\overline \sigma^4-h\overline \sigma\;. \ \
\eqa
 Stationarity condition for the effective potential (\ref{effpot}), gives 
\bqa
\label{para5}
h&=& m^2_\pi \overline{\sigma}
\eqa
The minimum of the effective potential at the tree level is given by $\overline \sigma=f_\pi$.Here it is pertinent to mention that when one reads the coefficient of $\overline {\sigma}^2 \over 2$ as the one single coefficient $(m^2-c)$ in which $m_\eta$ dependence cancles out and the coefficient of $\overline {\sigma}^4 \over 4$ as the another single coefficient  $(\lambda_1+\frac{\lambda_2}{2})$ where the $m_\eta$ and $m_{a_0}$ dependence cancles out, we see that the tree level effective potential of the $ SU(2)_{L} \times SU(2)_{R} $ sigma model becomes equivalent to the tree level potential of the $O(4)$ sigma model,where the degrees of freedom are the $\sigma$ and $\pi$ only.

We are considering a spatially uniform thermodynamic system in the equilibrium at temperature $T$ and chemical potentail $\mu_q \ (q=u \  \text{and} \  d)$.The partition function is written as the path integral over the quark/antiquark and meson fields \cite{scav,Schaefer:09,TiPQM3F} ,

\bqa
\nonumber
\mathcal{Z}&=&{\rm Tr}\exp\left[-\beta\left(\hat{\mathcal{H}}-\mu \hat{\mathcal{N}}\right)\right]\;\\ 
\nonumber
&=&\int \mathcal{D}\sigma\mathcal{D}\vec{a_0}\mathcal{D}\eta\mathcal{D}\vec{\pi}\int \mathcal{D}\psi\mathcal{D}\bar{\psi}\exp\left[-\int^{\beta}_0d\tau\int_Vd^3x \right. \\ &&\left.\left(\mathcal{L}^\mathcal{E}_{\mathcal{QM}}+\sum_{q=u,d}\mu\bar{q}\gamma^0q\right)\right]
\eqa
Where $V$ is the volume of the system, $\beta=\frac{1}{T}$, and the superscript $\mathcal{E}$ denotes the Euclidean Lagrangian.In this paper,we assume that the masses of $u$ and $d$ quarks are equal in  magnitude.Thus the quark chemical potential of the $u$ and $d$ quarks become equal,i.e. $\mu=\mu_u=\mu_d$.We evaluate the partition function in the mean field approximation.We replace the meson fields by their expectation value $<\mathcal{M}>=t_0\overline{\sigma}$, and neglect both the thermal and quantum fluctuations of the meson fields,while the quarks and antiquarks are retained as quantum fields.

In the mean-field approximation, the thermodynamic grand
potential for the QM model is given as

\bqa
  \nonumber
  \Omega_{\rm MF}(T,\mu;\overline {\sigma})&=&-T\frac{\ln\mathcal{Z}}{V}\; \\ 
  &=&U(\overline {\sigma} ) +
  \Omega_{q\bar{q}} (T,\mu;\overline {\sigma}). 
\label{Omega_MF}
\eqa

The quark/antiquark  contribution is given by

\bqa
\label{vac1}
\Omega_{q\bar{q}} (T,\mu;\overline{\sigma}) &=& \Omega_{q\bar{q}}^{vac}+\Omega_{q\bar{q}}^{T,\mu}\;\\
\Omega_{q\bar{q}}^{vac} &=&- 2 N_c\sum_q  \int \frac{d^3 p}{(2\pi)^3} E_q \theta( \Lambda_c^2 - \vec{p}^{2})\;\\
\label{vac2}
\nonumber
\Omega_{q\bar{q}}^{T,\mu}&=&- 2 N_c\sum_q \int \frac{d^3 p}{(2\pi)^3} T \left[ \ln \left(1+e^{-E_{q}^{+}/T}\right)\right. \\ &&\left.+\ln\left(1+e^{-E_{q}^{-}/T}\right)\right]\;
\label{vac3}
\eqa
The first term of the Eq.~(\ref{vac1}) denotes the fermion vacuum
contribution,where $\Lambda_c$ is the ultraviolet cutoff. E$_{q}^{\pm} =E_q \mp \mu $ and $E_q=\sqrt{p^2 + m{_q}{^2}}$ is the flavor dependent single particle energy of quark/antiquark, $m_q=\dfrac{g \overline{\sigma}}{2}$ is the mass of the given quark flavor.

Neglecting the quark one loop vacuum term of Eq.~(\ref{vac1}) in the the standard mean-field approximation (s-MFA),the QM model grand potential is written as,
\bqa
\nonumber
 \Omega_{QM}((\overline{\sigma},T,\mu))&=&U(\overline {\sigma} )+\Omega_{q\bar{q}}^{T,\mu}
 \eqa
\begin{equation}
\frac{\partial \Omega_{QM}(\overline{\sigma},T,\mu)}{\partial
      \overline{\sigma}} =0.
\label{EoMMF1}
\end {equation}
The global minima of the grand potential in the Eq.(\ref{EoMMF1}),gives the chiral condensate as a function of the temperature T and chemical potential $\mu$.

\section{QM model with Vacuum Term}
We will describe the calculation of the effective potential when the quark one-loop vacuum divergence of Eq.~(\ref{vac1}) is properly regularized using the minimal subtraction scheme and the
$\sigma$ and $\pi$ meson curvature mass (screening mass),defined by the second derivative of the thermodynamic potential at its minimum, has been used for fixing the model parameters.The zero temperature quark one-loop vacuum contribution can be written as 

\bqa
\nonumber
\label{vacqq}
\Omega^{vac}_{q\bar q}&=&-2N_c\sum_q\int \frac{d^3p}{(2\pi)^3}E_q\;\\
&&=-2N_c\sum_q \int \frac{d^4p}{(2\pi)^4}\ln(p^2_0+E^2_q)+K
\eqa

The infinite constant $K$ is dropped.When the Eq.(\ref{vacqq}) is dimensionally regularized near three dimensions, $d=3-2\epsilon$, one gets the $\epsilon$ zeroth order potential as :
\bqa
\label{omegavacc}
\nonumber
\Omega^{vac}_{q\bar q}&=&\frac{N_c}{(4 \pi)^2}\sum_q m^4_q\left[\frac{1}{\epsilon}+\frac{3}{2}+\ln(4\pi e^{-\gamma_E})+\ln\left(\Lambda^2\over m^2_q \right)\right]\;\\
\eqa
Redefining $\Lambda^2\longrightarrow \Lambda^2\frac{e^{\gamma_E}}{4\pi}$  in Eq.(\ref{omegavacc}), one gets
\bqa
\Omega^{vac}_{q\bar q}&=&\frac{N_c}{(4 \pi)^2}\sum_q m^4_q\left[\frac{1}{\epsilon}+\frac{3}{2}+\ln\left(\Lambda^2\over m^2_q \right)\right]\;
\eqa
Where $\Lambda$ is renormalization scale.

The thermodynamic potential is renormalized  by adding the following counterterm to the Lagrangian of the QM model.
\bqa
\delta \mathcal{L}&=&\frac{N_c}{(4 \pi)^2}\sum_q \frac{m^4_q}{\epsilon}\;.
\eqa
Now the first term of Eq.(\ref{vac1}) is replaced by the appropriately renormalized quark one-loop vacuum contribution of Eq.(\ref{omegavacc})
\bqa
\label{omegavacapp}
\Omega^{vac}_{q\bar q}&=&\frac{N_c}{(4 \pi)^2}\sum_q m^4_q\left[\frac{3}{2}+\ln\left(\Lambda^2\over m^2_q \right)\right]\;
\eqa
Since the vacuum ($\mu=0$ and $T=0$) grand potential gets contributions from $U(\overline{\sigma})$ and $\Omega^{vac}_{q\bar q} $, it becomes renormalization scale dependent.
\bqa
\label{qmvtomegavac}
\Omega^{\Lambda}(\overline{\sigma})&=&U(\overline{\sigma})+\Omega^{vac}_{q\bar q}\;.
\eqa
The unknown model parameters $m^2$,$c$,$\lambda_1$,$\lambda_2$,$h$ are obtained using the meson curvature masses,which are found by taking the second derivative of Eq.(\ref{qmvtomegavac}) evaluated at the minimum with respect to the different meson fields.The details of the model parameter determination, are presented in the Appendix (\ref{appenA}).

When the calculated model parameters get substituted in the expression of $U(\overline{\sigma})$,one can  rewrite the Eq.(\ref{qmvtomegavac}) as: 
\bqa
\nonumber
\Omega^{\Lambda}(\overline{\sigma})&=&\frac{1}{2}\left(m^2_{s}-\dfrac{N_cg^4f^2_\pi}{2(4\pi)^2}\right)\overline{\sigma}^2-\frac{1}{2}c\overline{\sigma}^2+\frac{1}{4}\left(\lambda_1+\frac{\lambda_{2s}}{2} \right.\\ \nonumber
&&\left.-\frac{N_cg^4}{2(4\pi)^2}\ln\left(\frac{4\Lambda^2}{g^2f^2_\pi}\right)\right)\overline{\sigma}^4-h\overline{\sigma}\\
&&+\frac{N_cg^4\overline{\sigma}^4}{8(4 \pi)^2}\left[\frac{3}{2}+\ln\left(\Lambda^2\over m^2_q \right)\right]\;,
\eqa
After the rearrangement of terms,we find the complete cancellation of renormalization scale $\Lambda$ in the vacuum grand potential.It is recast as
\bqa
\nonumber
\Omega(\overline{\sigma})&=&\frac{1}{2}\left(m^2_{s}-\dfrac{N_cg^4f^2_\pi}{2(4\pi)^2}\right)\overline{\sigma}^2-\frac{1}{2}c\overline{\sigma}^2+\frac{1}{4}\left(\lambda_1+\frac{\lambda_{2s}}{2}\right. \\
&&\left.+\dfrac{3N_cg^4}{4(4\pi)^2}\right)\overline{\sigma}^4-h\overline{\sigma}+\frac{N_cg^4\overline{\sigma}^4}{8(4\pi)^2}\ln\left(\dfrac{f^2_\pi}{\overline{\sigma}^2}\right)
\eqa
Now,in the presence of appropriately renormalized quark one-loop vacuum contribution,the thermodynamic grand potential of the  quark meson model with vacuum term (QMVT) will be written as :

\bqa
\label{qmvtomega}
\Omega_{\rm QMVT}((\overline{\sigma},T,\mu))&=&\Omega(\overline{\sigma})+\Omega^{T,\mu}_{q\bar q}
\eqa
\begin{equation}
\frac{\partial \Omega_{QMVT}(\overline{\sigma},T,\mu)}{\partial
      \overline{\sigma}} =0.
\label{EoMMF2}
\end {equation}
The global minima of the
grand potential in the Eq.(\ref{EoMMF2}),gives the PQMVT model chiral condensate $\overline{\sigma}$ 
as a function of the temperature $T$ and chemical potential $\mu$.
\section{Renormalized Quark Meson Model}
Several of the relatively recent investigations in the above detailed QMVT model framework have used  the
standard procedure of equating the vacuum expectation value of the sigma 
mean field to the pion decay constant and then putting the 
$\sigma,\ \vec{a_{0}}, \  \vec{\pi}$ and $\eta $ meson masses equal to
the their curvature (or screening) masses\cite{lars,guptiw,schafwag12,chatmoh1,TranAnd,vkkr12,chatmoh2,vkkt13,Herbst,Weyrich,kovacs,zacchi1,zacchi2,Rai}.However,in principle the physical masses of the mesons are given by the pole of 
their propagators and the residue of the pion propagator at its pole is related to 
the pion decay constant \cite{BubaCar,Naylor,fix1}.Furthermore,
since the effective potential is the generator of the n-point functions of the theory at zero external momenta,the curvature masses are equivalent to defining the meson masses using the evaluation of self-energy at zero momentum \cite{laine,Adhiand1,Adhiand2,Adhiand3}.It has been emphasized that the pole definition is the physical and gauge invariant one \cite{Kobes,Rebhan}.The curvature mass prescription is equivalent
to the pole mass prescription for the parameter fixing of the model in absence of Dirac sea contributions
but when the one quark loop vacuum correction is incorporated, the screening masses of mesons start to differ from the pole masses \cite{BubaCar,fix1}.In view of the above considerations,it becomes necessary to use the following detailed description of the exact on-shell parameter fixing procedure for the renormalized quark-meson (RQM) model where physical (pole) masses of the mesons and pion decay constant,are put into the relation of the running mass parameter and couplings by using the on-shell and the minimal subtraction renormalization schemes\cite{Adhiand2,asmuAnd}. 

\subsection{Self-Energies and Counterterms}
\begin{figure*}[htb]
\subfigure[\ One loop self-energy diagrams for sigma particle.]{
\label{sigmen} 
\begin{minipage}[b]{0.48\textwidth}
\centering \includegraphics[width=\linewidth]{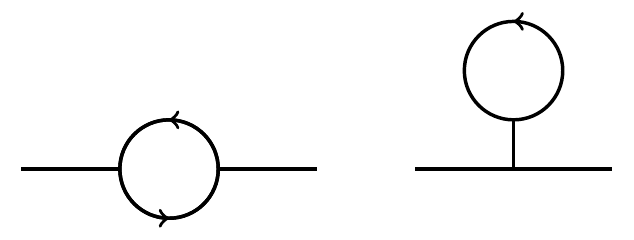}
\end{minipage}}%
\hfill
\subfigure[\ One loop self-energy diagrams for the $a_0$]{
\label{a0} 
\begin{minipage}[b]{0.48\textwidth}
\centering \includegraphics[width=\linewidth]{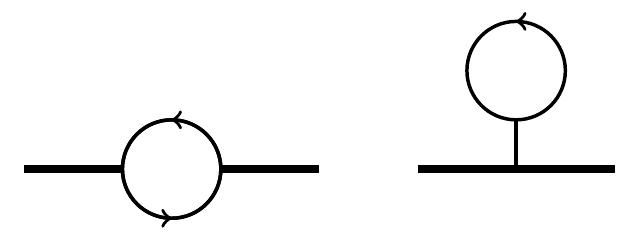}
\end{minipage}}
\caption{}
\label{fig:mini:fig1} 
\end{figure*}
\begin{figure*}[htb]
\subfigure[\ One-loop self-energy diagrams for the eta.]{
\label{eta} 
\begin{minipage}[b]{0.48\textwidth}
\centering \includegraphics[width=\linewidth]{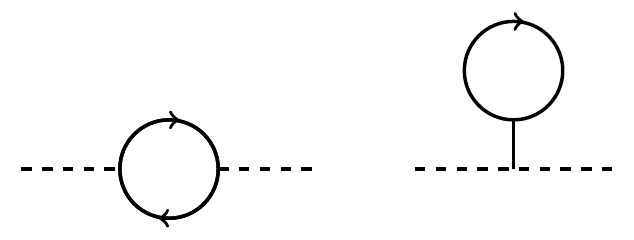}
\end{minipage}}%
\hfill
\subfigure[\ One-loop self-energy diagrams for the pion.]{
\label{pion} 
\begin{minipage}[b]{0.48\textwidth}
\centering \includegraphics[width=\linewidth]{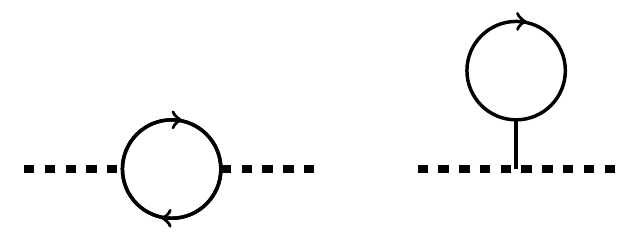}
\end{minipage}}
\caption{}
\label{fig:mini:fig1} 
\end{figure*} 

Once quark one-loop corrections are taken into account,
the tree level parameters of Eqs. (\ref{para1})--(\ref{para4}) 
become inconsistent unless the on-shell renormalization 
prescription is used.Though the dimensional regularization is used
to regularize the divergent loop integrals in the on-shell scheme,
the counterterm choices  are different from the minimal subtraction 
scheme.The loop corrections to the self-energies are cancelled exactly 
by the suitable choice of counterterms in the on-shell scheme.The 
renormalized parameters become  renormalization scale independent
as couplings are evaluated on shell.The wave functions/fields and parameters of Eq.(\ref{lag}) are bare quantities.The counterterms $\delta m^{2} $, $\delta g^{2} $, $\delta \lambda_{1} $,  $\delta \lambda_{2} $, $\delta c$ and $\delta h$ for the parameters and the counterterms $\delta Z_\sigma $,$\delta Z_{a_0} $,$\delta Z_\eta $,$\delta Z_\pi $, $\delta Z_\psi $ and $\delta Z_{\overline{\sigma}} $ for the wave functions/fields are introduced in the Lagrangian (\ref{lag}) where the renormalized fields and couplings are defined as
\bqa
\label{ctrm1}
\sigma_b=\sqrt{Z_\sigma}\sigma, \ \eta_b=\sqrt{Z_\eta}\eta, \ a^i_{0b}=\sqrt{Z_{a_{0}}}a_{0}
\\ \pi^i_b=\sqrt{Z_\pi}\pi, \ \psi_b=\sqrt{Z_\psi}\psi, \ m^2_b=Z_m m^2 \\
\lambda_{1b}=Z_{\lambda_1} \lambda_1, \ \lambda_{2b}=Z_{\lambda_2} \lambda_2, \ g_b=\sqrt{Z_g}g
\\ h_b=Z_h h,\ \ \ c_b=Z_c c,\ \overline{\sigma}_b=\sqrt{Z_{\overline{\sigma}}}\overline{\sigma}
\label{ctrm5}
\eqa
Where the $Z_ {(\sigma,a_0,\eta,\pi,\psi,\overline{\sigma} )}=1+\delta Z_{(\sigma,a_0,\eta,\pi,\psi,\overline{\sigma} )} $, denote  the field strength renormalization constant while $Z_ {(m,\lambda_1,\lambda_2,g,h,c )}=1+\delta Z_{(m,\lambda_1,\lambda_2,g,h,c )} $ denote the mass and coupling renormalization constant.Following the Ref.\cite{Adhiand1,Adhiand2,Adhiand3,asmuAnd},counterterms $\delta m^{2} $,$\delta \lambda_{1} $,  $\delta \lambda_{2} $, $\delta c$ and $\delta g^{2} $, $\delta h$ can be expressed in terms of the counterterms $\delta m^{2}_{\sigma} $,$\delta m^{2}_{a_0} $,$\delta m^{2}_{\eta} $, $\delta m^{2}_{\pi} $ and  $\delta m_{q} $, $\delta \overline \sigma^2 $.Using Eqs.(\ref{m1})--(\ref{m4}) together with Eqs. (\ref{ctrm1})-(\ref{ctrm5}), we can write
\bqa
\label{delta:lambda_1}
\delta \lambda_{1}&=&\frac{\delta m^2_\sigma+\delta m^2_\eta-\delta m^2_{a_0}-\delta m^2_\pi}{2 \ \overline \sigma^2 }-\lambda_1 \frac{\delta  \overline \sigma^2}{ \overline \sigma^2}
\\
\label{delta:lambda_2}
\delta \lambda_{2}&=&\frac{\delta m^2_{a_0}-\delta m^2_\eta}{\overline \sigma^2}-\lambda_2 \frac{\delta  \overline \sigma^2}{ \overline \sigma^2}
\\
\delta c&=&\frac{\delta m^2_\eta-\delta m^2_\pi}{2}
\\
\delta m^2&=&\delta m^2_\pi+\frac{\delta m^2_\eta-\delta m^2_\sigma}{2}
\\
\frac{\delta g^2}{4}&=&\frac{\delta m^2_q}{\overline \sigma^2}-g^2\frac{\delta  \overline \sigma^2}{4 \ \overline \sigma^2}
\label{delta:g}
\eqa
The one-loop correction at the pion-quark vertex is of order $N_{c}^0$.Hence $Z_\psi=1$ and the quark self energy correction $\delta m_q=0$ at this order.In consequence,we get $\sqrt{Z_\psi} \ \sqrt{Z_{g^2} \ g^2}=1$, and $\frac{\delta g^2}{g^2} \ + \delta Z_\pi=0$.Furthermore the $\delta m_q=0$ implies that $\delta g \ \overline \sigma/2 + g \ \delta \overline \sigma/2 =0 $.Eq.(\ref{delta:g}) gives 
\bqa
\label{Zpi}
\frac{\delta  \overline \sigma^2}{ \overline \sigma^2}&=&-\frac{\delta g^2}{g^2}=\delta Z_\pi
\eqa
Now one can rewrite Eq.(\ref{delta:lambda_1}),(\ref{delta:lambda_2}) as
\bqa
\label{delta:lambda_1n}
\delta \lambda_{1}&=&\frac{\delta m^2_\sigma+\delta m^2_\eta-\delta m^2_{a_0}-\delta m^2_\pi}{2 \ \overline \sigma^2}-\lambda_1 \delta Z_\pi
\\
\label{delta:lambda_2n}
\delta \lambda_{2}&=&\frac{\delta m^2_{a_0}-\delta m^2_\eta}{\overline \sigma^2}-\lambda_2 \delta Z_\pi
\eqa

The Feynman diagrams for the meson self-energies are drawn in the Figs.The scalar $\sigma$  and $a_0$ mesons are represented  by a solid line and a thick solid line respectively in the Fig.\ref{sigmen} and Fig.\ref{a0} where an arrow on the solid line denotes a quark.The corresponding self energy expressions for the $\sigma$ and $a^i_0$ are written as
\bqa
\nonumber
\label{selfeng1}
\Sigma_\sigma(p^2)&=&-\frac{2N_cg^2}{(4\pi)^2}\left[\mathcal{A}(m_q^2)-\mbox{$1\over2$}(p^2-4 m^2_q)\mathcal{B}(p^2)\right]
\\ &&
+{24(\lambda_1+\mbox{$\lambda_2\over 2$}) g \overline{\sigma}N_cm_q\over m_{\sigma}^2}\mathcal{A}(m_q^2)\;,
\\
\nonumber
\label{selfeng2}
\Sigma_{a_0}(p^2)&=&-\frac{2N_cg^2}{(4\pi)^2}\left[\mathcal{A}(m_q^2)-\mbox{$1\over2$}(p^2-4 m^2_q)\mathcal{B}(p^2)\right]
\\ &&
+{8(\lambda_1+\mbox{$3\lambda_2\over 2$}) g \overline{\sigma}N_cm_q\over m_{\sigma}^2}\mathcal{A}(m_q^2)\;,
\eqa
The pseudo-scalar $\eta$ and $\pi^i_0$ mesons in the Fig.\ref{eta} and the Fig.\ref{pion} are drawn by a dashed line and a thick dashed line in respective order.The corresponding self-energy expressions for the   $\eta$ and $\pi^i_0$ are given by

\bqa
\nonumber
\label{selfeng3}
\Sigma_\eta(p^2)&=&-\frac{2N_cg^2}{(4\pi)^2}\left[\mathcal{A}(m_q^2)-\mbox{$1\over2$}p^2\mathcal{B}(p^2)\right]
\\ &&
+{8(\lambda_1+\mbox{$\lambda_2\over 2$}) g \overline{\sigma}N_cm_q\over m_{\sigma}^2}\mathcal{A}(m_q^2)\;.
\\
\label{selfeng4}
\nonumber
\Sigma_\pi(p^2)&=&-\frac{2N_cg^2}{(4\pi)^2}\left[\mathcal{A}(m_q^2)-\mbox{$1\over2$}p^2\mathcal{B}(p^2)\right]
\\ &&
+{8(\lambda_1+\mbox{$\lambda_2\over 2$}) g \overline{\sigma}N_cm_q\over m_{\sigma}^2}\mathcal{A}(m_q^2)\;,
\eqa

\begin{figure}[htb]
\begin{center}
\includegraphics[width=0.45\textwidth]{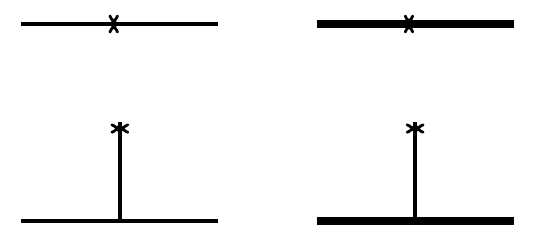}
\end{center}
\caption{Counterterm for the two-point functions of the scalar $\bf{\sigma}$ and $\bf{a_0}$ meson}.
\label{count11}
\end{figure}

\begin{figure}[htb]
\begin{center}
\includegraphics[width=0.45\textwidth]{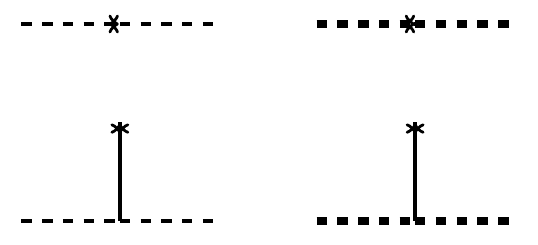}
\end{center}
\caption{Counterterm for the two-point functions of the pseudo-scalar $\eta$ 
and $\pi$ meson.}
\label{count22}
\end{figure}
Where the $\mathcal{A}(m^2_q)$ and $\mathcal{B}(p^2)$ are defined in the Appendix (\ref{appenB}) and the last term of the Eqs.(\ref{selfeng1}), (\ref{selfeng2}), (\ref{selfeng3}) and (\ref{selfeng4}) are the contributions of the tadpole diagrams to the self energies.The counterterm diagrams for the two-point functions of the scalar mesons $ \bf{\sigma} $, $\bf{a_0}$ and the pseudo-scalar mesons $\bf{\eta}$, $\bf{\pi}$ are shown respectively in the Fig.\ref{count11} and the Fig.\ref{count22}.

\begin{figure}[htb]
\begin{center}
\includegraphics[width=0.45\textwidth]{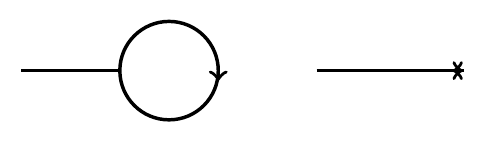}
\end{center}
\caption{One point diagram for the sigma particle and its counterterm. }
\label{tad}
\end{figure}

The diagram for the quark one-loop correction to the one-point function and its counterterm diagram, is shown in Fig. \ref{tad}. It can be written as 
\bqa
\delta\Gamma^{(1)}
&=&-
4 N_c g m_q \mathcal{A}(m_q^2)+i\delta t
\;,
\eqa

\subsection{Parameters with  Renormalization}
The vanishing of the one-point function $\Gamma^{(1)}=it=i(h-m_{\pi}^2 \  \overline \sigma)$ gives the tree level equation of motion $t=0$ and fixes the classical minimum of the effective potential.The first renormalization condition $<\sigma>=0$
requires that the one loop correction $\delta \Gamma^{(1)}$ to the one 
point function is put to zero such that the minimum of the effective potential does not change.Thus the first renormalization condition $\delta\Gamma^{(1)}=0 $ gives 
\bqa
\delta t
&=&-4i N_c g m_q \mathcal{A}(m_q^2)  \;.
\eqa
The equation $h=t+m_{\pi}^2 \  \overline \sigma $ enables the writing of the
counterterm $\delta h$ in terms of the tadpole counterterm  $\delta t$
\bqa
\label{delta:h}
\delta h&=&m^2_\pi \ \delta \overline \sigma +\delta m^2_\pi \ \overline \sigma +\delta t
\eqa
Using Eq.(\ref{Zpi}) we can write
\bqa
\label{delta:hn}
\delta h&=&\frac{1}{2}m^2_\pi \ \overline \sigma  \delta Z_\pi+\delta m^2_\pi \ \overline \sigma +\delta t
\eqa
We can write the inverse propagator for the scalar $ \sigma,a_0 $ and pseudo-scalar
$\pi,\eta $ mesons as 
\bqa
p^2-m_{\sigma,a_0,\pi,\eta}^2-i\Sigma_{\sigma,a_0,\pi,\eta}(p^2)
{\rm +counterterms}
\;.
\label{definv}
\eqa

The renormalized mass in the Lagrangian is put equal to the physical mass, i.e.\ $m=m_{\rm pole}$\footnote{The contribution of the imaginary parts of the self-energies into defining the mass has not been considered.} when the on-shell scheme gets implemented and we can write

\bqa
\Sigma(p^2=m_{\sigma,a_0,\pi,\eta}^2)
{\rm +counterterms}
&=&0
\label{pole}
\;.
\eqa
Since the propagator residue is put to unity in the on shell scheme, one gets 
\bqa
\label{res}
{\partial\over\partial p^2}\Sigma_{\sigma,a_0,\pi,\eta}(p^2)\Big|_{p^2=m_{\sigma,a_0,\pi,\eta}^2}
{\rm +counterterms}
&=&0\;.
\eqa

The Fig. \ref{count11} and Fig. \ref{count22} diagrams for the counterterms of the two point functions of the  scalar $\sigma,a_0$ and pseudo-scalar $\pi,\eta$ mesons, are written as 
\bqa
\label{count1}
\Sigma_{\sigma}^{\rm ct1}(p^2)&=&i\left[\delta Z_{\sigma}(p^2-m_{\sigma}^2)-\delta m_{\sigma}^2\right]\;,
\eqa
\bqa
\label{count2}
\Sigma_{a_0}^{\rm ct1}(p^2)&=&i\left[\delta Z_{a_0}(p^2-m_{a_0}^2)-\delta m_{a_0}^2\right]\;,
\eqa
\bqa
\label{count3}
\Sigma_{\pi}^{\rm ct1}(p^2)&=&i\left[\delta Z_{\pi}(p^2-m_{\pi}^2)-\delta m_{\pi}^2\right]\;,
\eqa
\bqa
\label{count4}
\Sigma_{\eta}^{\rm ct1}(p^2)&=&i\left[\delta Z_{\eta}(p^2-m_{\eta}^2)-\delta m_{\eta}^2\right]\;,
\eqa

\bqa
\label{count5}
\Sigma^{\rm ct2}_{\sigma}&=&3\Sigma^{\rm ct2}_{\pi}=3\Sigma^{\rm ct2}_{\eta}\;,
\\
&=&-{24(\lambda_1+\mbox{$\lambda_2\over 2$}) g \overline{\sigma}N_cm_q\over m_{\sigma}^2}\mathcal{A}(m_q^2)\;,
\\
\Sigma^{\rm ct2}_{a_0}&=&-{8(\lambda_1+\mbox{$3\lambda_2\over 2$}) g \overline{\sigma}N_cm_q\over m_{\sigma}^2}\mathcal{A}(m_q^2)\;,
\eqa

The counterterms of Eq.(\ref{count5}) completely cancel the respective tadpole contributions to  the self-energies of $\sigma, a_0$ and $\pi,\eta$.The evaluation of the self-energies and their derivatives in on-shell conditions give all the  renormalization constants.When Eqs. (\ref{pole}),(\ref{res}) and (\ref{count1})--(\ref{count4}) are combined,we obtain the following set of equations.

\bqa
\delta m_{\sigma}^2&=&-i\Sigma_{\sigma}(m_{\sigma}^2)\;;\delta Z_\sigma =
i{\partial\over\partial p^2}\Sigma_\sigma(p^2)\Big|_{p^2=m_\sigma^2}\;,
\eqa
\bqa
\delta m_{a_0}^2&=&-i\Sigma_{a_0}(m_{a_0}^2)\;;\delta Z_{a_0} =
i{\partial\over\partial p^2}\Sigma_{a_0}(p^2)\Big|_{p^2=m_{a_0}^2}\;,
\eqa
\bqa
\delta m_{\pi}^2&=&-i\Sigma_{\pi}(m_{\pi}^2)\;;\delta Z_\pi= i{\partial\over\partial p^2}\Sigma_\pi(p^2)\Big|_{p^2=m_\pi^2}\;,
\eqa
\bqa
\delta m_{\eta}^2&=&-i\Sigma_{\eta}(m_{\eta}^2)\;;\delta Z_\eta =i{\partial\over\partial p^2}\Sigma_\eta(p^2)\Big|_{p^2=m_\eta^2}\;.
\eqa
When the self energy (neglecting the tadpole contributions) expressions from the Eqs. (\ref{selfeng1})--(\ref{selfeng4}) are used, we get the following equations.
\bqa
\label{countf1}
\delta m_{\sigma}^2
&=&
2ig^2N_c\left[\mathcal{A}(m_q^2)-\mbox{$1\over2$}(m_{\sigma}^2-4m_q^2)\mathcal{B}(m_{\sigma}^2)
\right]\;,
\eqa
\bqa
\delta m_{a_0}^2
&=&
2ig^2N_c\left[\mathcal{A}(m_q^2)-\mbox{$1\over2$}(m_{a_0}^2-4m_q^2)\mathcal{B}(m_{a_0}^2)
\right]\;,
\eqa
\bqa
\delta m_{\pi}^2
&=&
2ig^2N_c\left[\mathcal{A}(m_q^2)-\mbox{$1\over2$}m_{\pi}^2\mathcal{B}(m_{\pi}^2)
\right]\;,
\eqa
\bqa
\delta m_{\eta}^2
&=&
2ig^2N_c\left[\mathcal{A}(m_q^2)-\mbox{$1\over2$}m_{\eta}^2\mathcal{B}(m_{\eta}^2)
\right]\;,
\eqa
\bqa
\delta Z_{\sigma}&=&
ig^2N_c\left[\mathcal{B}(m_{\sigma}^2)+(m_{\sigma}^2-4m_q^2)\mathcal{B}^{\prime}(m_{\sigma}^2)
\right]\;,
\eqa
\bqa
\delta Z_{a_0}&=&
ig^2N_c\left[\mathcal{B}(m_{a_0}^2)+(m_{a_0}^2-4m_q^2)\mathcal{B}^{\prime}(m_{a_0}^2)
\right]\;,
\eqa
\bqa
\delta Z_{\pi}&=&
ig^2N_c\left[\mathcal{B}(m_{\pi}^2)+m_{\pi}^2\mathcal{B}^{\prime}(m_{\pi}^2)
\right]\;,
\eqa
\bqa
\label{countf8}
\delta Z_{\eta}&=&
ig^2N_c\left[\mathcal{B}(m_{\eta}^2)+m_{\eta}^2\mathcal{B}^{\prime}(m_{\eta}^2)
\right]\;.
\eqa

Exploiting the Eqs. (\ref{countf1})--(\ref{countf8}) together with the Eqs. (\ref{delta:lambda_1}) --(\ref{Zpi}) and (\ref{delta:hn}), we find the following expressions for the counterterms in the on-shell scheme.  

\begin{widetext}
 \bqa \nonumber
 \label{os1}
\delta\lambda_{1\os}&=&\frac{iN_cg^2}{\overline{\sigma}^2}\left[-\frac{1}{2}(m^2_\sigma-4m^2_q)\mathcal{B}(m^2_\sigma)-\frac{1}{2}m^2_\eta \mathcal{B}(m^2_\eta)+\frac{1}{2}(m^2_{a_0}-4m^2_q)\mathcal{B}(m^2_{a_0})+\frac{1}{2}m^2_\pi \mathcal{B}(m^2_\pi)\right]-\lambda_1ig^2N_c\left[\mathcal{B}(m_{\pi}^2)\right. \\ \nonumber
&&\left.+m_{\pi}^2\mathcal{B}^{\prime}(m_{\pi}^2)\right]\\
\nonumber
&=&\delta \lambda_{1\text{div}}+2\lambda_1\dfrac{N_cg^2}{(4\pi)^2}\ln\left(\frac{\Lambda^2}{m_q^2}\right)+\dfrac{N_cg^2}{(4\pi)^2}\left[\frac{(m^2_\sigma-4m^2_q)\mathcal{C}(m^2_\sigma)+m^2_\eta \mathcal{C}(m^2_\eta)-(m^2_{a_0}-4m^2_q)\mathcal{C}(m^2_{a_0})-m^2_\pi \mathcal{C}(m^2_\pi)}{2\overline{\sigma}^2}\right. \\
&&\left.+\lambda_1(\mathcal{C}(m^2_\pi)+m^2_\pi\mathcal{C}^\prime(m^2_\pi))\right]\;,
\\
\nonumber
\delta\lambda_{2\os} 
&=&\frac{iN_cg^2}{\overline{\sigma}^2}\left[-(m^2_{a_0}-4m^2_q)\mathcal{B}(m^2_{a_0})+m^2_\eta \mathcal{B}(m^2_\eta)\right]-\lambda_2ig^2N_c\left[\mathcal{B}(m_{\pi}^2)+m_{\pi}^2\mathcal{B}^{\prime}(m_{\pi}^2)\right]\;,
\\
\nonumber
&=&\delta \lambda_{2\text{div}}+\dfrac{N_cg^2}{(4\pi)^2}\left(2\lambda_2-g^2\right)\ln\left(\frac{\Lambda^2}{m_q^2}\right)\\
&&+\dfrac{N_cg^2}{(4\pi)^2}\left[\frac{(m^2_{a_0}-4m^2_q)\mathcal{C}(m^2_{a_0})-m^2_\eta \mathcal{C}(m^2_\eta)}{{\overline{\sigma}^2}}+\lambda_2(m^2_\pi \mathcal{C}^{\prime}(m^2_\pi)+\mathcal{C}(m^2_\pi))\right]\;,
\\
\nonumber
\delta m^2_{\os}&=&2iN_cg^2\left[\mathcal{A}(m^2_q)-\frac{1}{2}m^2_\pi \mathcal{B}(m^2_\pi)\right]+iN_cg^2\left[-\frac{1}{2}m^2_\eta \mathcal{B}(m^2_\eta)+\frac{1}{2}(m^2_\sigma-4m^2_q)\mathcal{B}(m^2_\sigma)\right]\;
\\
&=&\delta m^2_{\text{div}}+\dfrac{N_cg^2}{(4\pi)^2}m^2\ln\left(\frac{\Lambda^2}{m_q^2}\right)
+\dfrac{N_cg^2}{(4\pi)^2}\left[m^2_\pi \mathcal{C}(m^2_\pi)+\dfrac{m^2_{\eta}\mathcal{C}(m^2_{\eta})-(m^2_\sigma-4m^2_q) \mathcal{C}(m^2_\sigma)}{2}-2m^2_q\right]\;,
\\
\nonumber
\delta c_{\os}&=&\frac{iN_cg^2}{2}\left[-m^2_\eta \mathcal{B}(m^2_\eta)+m^2_\pi \mathcal{B}(m^2_\pi)\right]\;\\
&=&\delta c_{\text{div}}+\dfrac{N_cg^2}{(4\pi)^2}c\ln\left(\frac{\Lambda^2}{m_q^2}\right)+\dfrac{N_cg^2}{2(4\pi)^2}\left[m^2_{\eta}\mathcal{C}(m^2_{\eta})-m^2_\pi \mathcal{C}(m^2_\pi)\right]\;,
\\
\nonumber
\delta g^2_{\os}&=&-iN_cg^4\left[m^2_\pi \mathcal{B}^\prime(m^2_\pi)+\mathcal{B}(m^2_\pi)\right]\;\\
&=&\delta g^2_{\text{div}}+\dfrac{N_cg^4}{(4\pi)^2}\left[\ln\left(\frac{\Lambda^2}{m_q^2}\right)+\mathcal{C}(m^2_\pi)+m^2_\pi \mathcal{C}^{\prime}(m^2_\pi)\right]\;,
\\
\nonumber
\delta h_{\os}
&=&\dfrac{iN_cg^2}{2(4\pi)^2}h\left[m^2_\pi \mathcal{B}^\prime(m^2_\pi)-\mathcal{B}(m^2_\pi)\right]\;\\
&=&\delta h_{\text{div}}+\dfrac{N_cg^2}{2(4\pi)^2}h\left[\ln\left(\frac{\Lambda^2}{m_q^2}\right)+\mathcal{C}(m^2_\pi)-m^2_\pi \mathcal{C}^{\prime}(m^2_\pi)\right]\;\\
\nonumber
\label{os7}
\delta \overline{\sigma}^2_{\os}&=&iN_cg^2\overline{\sigma}^2\left[m^2_\pi \mathcal{B}^\prime(m^2_\pi)+\mathcal{B}(m^2_\pi)\right]\;\\
&=&\delta \overline{\sigma}^2_{\text{div}}-\dfrac{N_cg^2\overline{\sigma}^2}{(4\pi)^2}\left[\ln\left(\frac{\Lambda^2}{m_q^2}\right)+\mathcal{C}(m^2_\pi)+m^2_\pi \mathcal{C}^{\prime}(m^2_\pi)\right]\;\\
\delta Z^{\os}_{\sigma}&=&\delta Z_{\sigma,\rm div}-
\frac{N_cg^2}{(4\pi)^2}\left[\ln\left(\frac{\Lambda^2}{m_q^2}\right)+\mathcal{C}(m_{\sigma}^2)+(m_{\sigma}^2-4m_q^2)\mathcal{C}^{\prime}(m_{\sigma}^2)
\right]\;
\\
\delta Z^{\os}_{a_0}&=&\delta Z_{a_0,\rm div}-
\frac{N_cg^2}{(4\pi)^2}\left[\ln\left(\frac{\Lambda^2}{m_q^2}\right)+\mathcal{C}(m_{a_0}^2)+(m_{a_0}^2-4m^2_q)\mathcal{C}^{\prime}(m_{a_0}^2)
\right]\;
\\
\delta Z^{\os}_{\pi}&=&\delta Z_{\pi,\rm div}-
\frac{N_cg^2}{(4\pi)^2}\left[\ln\left(\frac{\Lambda^2}{m_q^2}\right)+\mathcal{C}(m_{\pi}^2)+m_{\pi}^2\mathcal{C}^{\prime}(m_{\pi}^2)
\right]\;
\\
\delta Z^{\os}_{\eta}&=&\delta Z_{\eta,\rm div}-
\frac{N_cg^2}{(4\pi)^2}\left[\ln\left(\frac{\Lambda^2}{m_q^2}\right)+\mathcal{C}(m_{\eta}^2)+m_{\eta}^2\mathcal{C}^{\prime}(m_{\eta}^2)
\right]\;
\eqa
here, $\lambda_1$, $\lambda_2$, $m^2$, $c$, $h$ and $g^2$ in Eqs.(\ref{os1})--(\ref{os7}) are the same as in the Eqs.(\ref{para1})--(\ref{para4}) and Eq.(\ref{para5}) for the QM model with the 'no Dirac sea' approximation.

The $\mathcal{B}(m^2),\mathcal{B}^{\prime}(m^2)$ and $\mathcal{C}(m^2),\mathcal{C}^{\prime}(m^2)$ are defined in the Appendix(\ref{appenB}).The divergent part of the counterterms are $\delta \lambda_{1\text{div}}=\frac{N_cg^22\lambda_1}{(4\pi)^2\epsilon}\;$, \ $ \delta \lambda_{2\text{div}}=\frac{N_cg^2}{(4\pi)^2\epsilon}(2\lambda_2-g^2)\;$, \ $\delta m^2_{\text{div}}=\frac{N_cg^2 m^2}{(4\pi)^2\epsilon}\;$, \ $\delta c_{\text{div}}=\frac{N_cg^2 c}{(4\pi)^2\epsilon}\;$, \ $\delta g^2_{\text{div}}=\frac{N_cg^4}{(4\pi)^2\epsilon}\; $, \ $\delta \overline{\sigma}^2_{\text{div}}=-\frac{N_cg^2\overline{\sigma}^2}{(4\pi)^2\epsilon}\; $, \ $ \delta h_{\text{div}}=\frac{N_cg^2 h}{2(4\pi)^2\epsilon}\;$, $\delta Z_{\sigma,\rm div}=\delta Z_{a_0,\rm div}=\delta Z_{\pi,\rm div}=\delta Z_{\eta,\rm div}=-\frac{N_cg^2}{(4\pi)^2\epsilon}$ .For both,the on-shell and the $\overline{\text{MS}}$ schemes, the divergent part of the counterterms are the same, i.e. $\delta \lambda_{1\text{div}}=\delta \lambda_{1\ms}$, \ $\delta \lambda_{2\text{div}}=\delta \lambda_{2\ms}$ etc.
\end{widetext}
Since the bare parameters are independent of the renormalization scheme, we can immediately write down the relations between the renormalized parameters in the on-shell and $\text{MS}$ schemes as the following
\bqa
\lambda_{1\ms}&=&\lambda_1+\delta \lambda_{1\os}-\delta \lambda_{1\ms}\;
\eqa
\bqa
\lambda_{2\ms}&=&\lambda_2+\delta \lambda_{2\os}-\delta \lambda_{2\ms}\;
\eqa
\bqa
m^2_{\ms}&=&m^2+\delta m^2_{\os}-\delta m^2_{\ms}\;
\eqa
\bqa
c_{\ms}&=&c+\delta c_{\os}-\delta c_{\ms}\;
\eqa
\bqa
h_{\ms}&=&h+\delta h_{\os}-\delta h_{\ms}\;
\eqa

\bqa
g^2_{\ms}&=&g^2+\delta g^2_{\os}-\delta g^2_{\ms}\;
\eqa
\bqa
\overline{\sigma}^2_{\ms}&=&\overline{\sigma}^2+\delta \overline{\sigma}^2_{\os}-\delta \overline{\sigma}^2_{\ms}\;.
\eqa
The minimum of the effective potential is at $\overline{\sigma}=f_\pi$ and the masses have the measured value in vacuum.Using the above set of equations together with the Eqs.(\ref{os1})--(\ref{os7}), we can write the scale $\Lambda$ dependent running parameters in the $\overline{\text{MS}}$ scheme as the following
\begin{widetext}
\bqa \nonumber
\label{params1}
\lambda_{1\ms}(\Lambda)&=&\lambda_1+2\lambda_1\dfrac{N_cg^2}{(4\pi)^2}\ln\left(\frac{\Lambda^2}{m_q^2}\right)+\dfrac{N_cg^2}{(4\pi)^2}\left[\frac{(m^2_\sigma-4m^2_q)\mathcal{C}(m^2_\sigma)+m^2_\eta \mathcal{C}(m^2_\eta)-(m^2_{a_0}-4m^2_q)\mathcal{C}(m^2_{a_0})-m^2_\pi \mathcal{C}(m^2_\pi)}{2f_\pi^2}\right. \\
&&\left.+\lambda_1(\mathcal{C}(m^2_\pi)+m^2_\pi\mathcal{C}^\prime(m^2_\pi))\right]\;,\\
\nonumber
\lambda_{2\ms}(\Lambda)
&=&\lambda_2+\dfrac{N_cg^2}{(4\pi)^2}\left(2\lambda_2-g^2\right)\ln\left(\frac{\Lambda^2}{m_q^2}\right)\\
&&+\dfrac{N_cg^2}{(4\pi)^2}\left[\frac{(m^2_{a_0}-4m^2_q)\mathcal{C}(m^2_{a_0})-m^2_\eta \mathcal{C}(m^2_\eta)}{{f_\pi^2}}+\lambda_2(m^2_\pi \mathcal{C}^{\prime}(m^2_\pi)+\mathcal{C}(m^2_\pi))\right]\;,
\\
m^2_{\ms}(\Lambda)
&=&m^2+\dfrac{N_cg^2}{(4\pi)^2}m^2\ln\left(\frac{\Lambda^2}{m_q^2}\right)
+\dfrac{N_cg^2}{(4\pi)^2}\left[m^2_\pi \mathcal{C}(m^2_\pi)+\dfrac{m^2_{\eta}\mathcal{C}(m^2_{\eta})-(m^2_\sigma-4m^2_q) \mathcal{C}(m^2_\sigma)}{2}-2m^2_q\right]\;,
\\
c_{\ms}(\Lambda)
&=&c+\dfrac{N_cg^2}{(4\pi)^2}c\ln\left(\frac{\Lambda^2}{m_q^2}\right)+\dfrac{N_cg^2}{2(4\pi)^2}\left[m^2_{\eta}\mathcal{C}(m^2_{\eta})-m^2_\pi \mathcal{C}(m^2_\pi)\right]\;,
\\
h_{\ms}(\Lambda)
&=&h+\dfrac{N_cg^2}{2(4\pi)^2}h\left[\ln\left(\frac{\Lambda^2}{m_q^2}\right)+\mathcal{C}(m^2_\pi)-m^2_\pi \mathcal{C}^{\prime}(m^2_\pi)\right]\;,
\\
\label{params6}
g^2_{\ms}(\Lambda)
&=&g^2+\dfrac{N_cg^4}{(4\pi)^2}\left[\ln\left(\frac{\Lambda^2}{m_q^2}\right)+\mathcal{C}(m^2_\pi)+m^2_\pi \mathcal{C}^{\prime}(m^2_\pi)\right]\;,
\\
\label{params7}
\overline{\sigma}^2_{\ms}(\Lambda)
&=&f_\pi^2-\dfrac{4N_cm^2_q}{(4\pi)^2}\left[\ln\left(\frac{\Lambda^2}{m_q^2}\right)+\mathcal{C}(m^2_\pi)+m^2_\pi \mathcal{C}^{\prime}(m^2_\pi)\right]\;.
\eqa
\end{widetext}
\begin{table*}[!htbp]
    \caption{Parameters of the QM model for $m_\sigma=400, 500 \ \text{and} \ 600 \ \text{MeV}$.}
    \label{tab:table2}
    \begin{tabular}{p{0.163\textwidth} p{0.163\textwidth}  p{0.163\textwidth} p{0.163\textwidth} p{0.163\textwidth} p{0.163\textwidth} }
      \toprule 
      $m_{\sigma}(\text{MeV})$& $c(\text{MeV}^2)$ & $m^2(\text{MeV}^2)$ & $\lambda_1$ & $\lambda_2$ & $h(\text{MeV}^3)$ \\
      \hline 
      \hline
      $400$ & $(374.28)^2$ & $(297.74)^2$ &$-30.61$ & $ 77.51$ & $(120.99)^3$  \\
     $500$ & $(374.28)^2$ & $(208.92)^2$ &$-23.32$ & $ 77.51$ & $(120.99)^3$  \\
     $600$ & $(374.28)^2$ & $-(106.54)^2$ &$-19.05$ & $ 77.51$ & $(120.99)^3$  \\
     \hline 
    \end{tabular}
\end{table*}
$\lambda_1$, $\lambda_2$, $m^2$, $c$, $h$ and $g^2$ in Eqs.(\ref{params1})--(\ref{params7}) have the same tree level QM model values that we obtain after putting $\overline{\sigma}=f_\pi$ in the Eqs.(\ref{para1})--(\ref{para4}) and Eq.(\ref{para5}).

In the large-$N_c$ limit the parameters $\lambda_{1\ms}$, $\lambda_{2\ms}$, $m^2_{\ms}$, $c_{\ms}$, $h_{\ms}$ and $g^2_{\ms}$ are running with the scale $\Lambda$ and satisfy a set of the following simultaneous renormalization group equations. 

\bqa
\label{diffpara1}
\dfrac{d\lambda_{1\ms}(\Lambda)}{d\ln(\Lambda)}&=&\dfrac{4N_c}{(4\pi)^2}g^2_{\ms}\lambda_{1\ms}\;,
\eqa
\bqa
\dfrac{d\lambda_{2\ms}(\Lambda)}{d\ln(\Lambda)}&=&\dfrac{2N_c}{(4\pi)^2}\left[2\lambda_{2\ms}g^2_{\ms}-g^4_{\ms}\right]\;,
\eqa
\bqa
\dfrac{dm^2_{\ms}(\Lambda)}{d\ln(\Lambda)}&=&\dfrac{2N_c}{(4\pi)^2}g^2_{\ms}m^2_{\ms}\;,
\eqa
\bqa
\dfrac{d c_{\ms}(\Lambda)}{d\ln(\Lambda)}&=&\dfrac{2N_c}{(4\pi)^2}g^2_{\ms}c_{\ms}\;,
\eqa
\bqa
\dfrac{d h_{\ms}(\Lambda)}{d\ln(\Lambda)}&=&\dfrac{N_c}{(4\pi)^2}g^2_{\ms}h_{\ms}\;,
\eqa
\bqa
\dfrac{d g^2_{\ms}}{d\ln(\Lambda)}&=&\dfrac{2N_c}{(4\pi)^2}g^4_{\ms}\;,
\eqa
\bqa
\label{diffpara7}
\dfrac{d \overline{\sigma}^2_{\ms}}{d\ln(\Lambda)}&=&-\dfrac{2N_c}{(4\pi)^2}g^2_{\ms}\overline{\sigma}^2_{\ms}
\eqa

Solving the differential Eqs.(\ref{diffpara1})--(\ref{diffpara7}), we get the following solutions
\bqa
\label{para01}
\lambda_{1\ms}(\Lambda)&=&\frac{\lambda_{10}}{\left(1-\dfrac{N_c g^2_0}{(4\pi)^2}\ln\left(\dfrac{\Lambda^2}{\Lambda_0^2}\right)\right)^2}\;,
\\
g^2_{\ms}(\Lambda)&=&\frac{g^2_0}{1-\dfrac{N_c g^2_0}{(4\pi)^2}\ln\left(\dfrac{\Lambda^2}{\Lambda^2_0}\right)}\;,
\\
\lambda_{2\ms}(\Lambda)&=&\frac{\lambda_{20}-\dfrac{N_c g^4_0}{(4\pi)^2}\ln\left(\dfrac{\Lambda^2}{\Lambda^2_0}\right)}{\left(1-\dfrac{N_c g^2_0}{(4\pi)^2}\ln\left(\dfrac{\Lambda^2}{\Lambda^2_0}\right)\right)^2}\;,  
\\
m^2_{\ms}(\Lambda)&=&\frac{m^2_0}{1-\dfrac{N_c g^2_0}{(4\pi)^2}\ln\left(\dfrac{\Lambda^2}{\Lambda^2_0}\right)}\;,
\\
c_{\ms}(\Lambda)&=&\frac{c_0}{1-\dfrac{N_c g^2_0}{(4\pi)^2}\ln\left(\dfrac{\Lambda^2}{\Lambda^2_0}\right)}\;,
\eqa
\bqa
\label{para06}
h_{\ms}(\Lambda)&=&\frac{h_0}{\sqrt{1-\dfrac{N_c g^2_0}{(4\pi)^2}\ln\left(\dfrac{\Lambda^2}{\Lambda^2_0}\right)}}\;,
\\
\label{para07}
\overline{\sigma}^2&=&f^2_\pi\left[1-\frac{N_cg^2_0}{(4\pi)^2}\ln\left(\frac{\Lambda^2}{\Lambda^2_0}\right)\right]\;
\eqa

Where the parameters $\lambda_{10}$,\ $\lambda_{20}$,\ $g^2_0$,\ $m^2_0$,\ $c_0$ and $h_0$, are the running parameter values at the scale $\Lambda_0$.We can choose the $\Lambda_0$ to satisfy the following relation
\bqa
\ln\left(\frac{\Lambda^2_0}{m_q^2}\right)+\mathcal{C}(m^2_\pi)+m^2_\pi \mathcal{C}^{\prime}(m^2_\pi)&=&0\;.
\eqa
Now, we can calculate the parameters of Eqs.(\ref{params1})--(\ref{params7}) at the scale $\Lambda=\Lambda_0$ and find $\lambda_{10}$,\ $\lambda_{20}$,\ $g^2_0$,\ $m^2_0$,\ $c_0$ and $h_0$.
\\ 

\subsection{Effective Potential}

Using the values of the parameters from the Eqs.(\ref{para01})--(\ref{para07}), the vacuum effective potential in the $\overline{\text{MS}}$ scheme can be written as

\bqa
\label{omegarqm}
\Omega_{vac}&=&U(\overline{\sigma}_{\ms})+\Omega^{q,vac}_{\ms}+\delta U(\overline{\sigma}_{\ms})
\eqa
Where
\bqa
\label{omegams1}
\nonumber
U(\overline{\sigma}_{\ms})&=&\frac{m_{\ms}^2(\Lambda)}{2}\overline{\sigma}^2_{\ms}-\frac{c_{\ms}(\Lambda)}{2}\overline{\sigma}^2_{\ms}\;\\
\nonumber
&&+\frac{1}{4}\left(\lambda_{1\ms}(\Lambda)+\frac{\lambda_{2\ms}(\Lambda)}{2}\right)\overline{\sigma}^4_{\ms}\\
&&-h_{\ms}(\Lambda)\overline{\sigma}_{\ms}\;,
\eqa

\begin{widetext}
\bqa
\nonumber
\delta U(\overline{\sigma}_{\ms})&=&\frac{1}{2}(\delta m^2_{\ms}-\delta c_{\ms})\overline{\sigma}_{\ms}^2+\frac{1}{2}(m^2_{\ms}-c_{\ms})\delta \overline{\sigma}^2_{\ms}+\frac{1}{4}\left(\delta \lambda_{1\ms}+\frac{\delta\lambda_{2\ms}}{2}\right) \overline{\sigma}^4_{\ms}+\frac{1}{4}\left(\lambda_{1\ms}+\frac{\lambda_{2\ms}}{2}\right)\delta \overline{\sigma}^4_{\ms}\; \\
&&-\delta h_{\ms} \overline{\sigma}_{\ms}-h_{\ms}\delta\overline{\sigma}_{\ms}\;
\eqa

The order $\mathcal{O}(N^2_c)$ terms are dropped as these are two loop terms and we get
\bqa
\delta U(\overline{\sigma}_{\ms})&=&-\frac{N_cg^4_{\ms}\overline{\sigma}^4_{\ms}}{8(4\pi)^2}\frac{1}{\epsilon}=-\frac{2N_c\Delta^4}{(4\pi)^2}\frac{1}{\epsilon}
\eqa

\bqa
\label{omegavac}
\Omega^{q,vac}_{\ms}&=&\frac{N_cg^4_{\ms}\overline{\sigma}^4_{\ms}}{8(4\pi)^2}\left[\frac{1}{\epsilon}+\frac{3}{2}+\ln\left(\frac{4\Lambda^2}{g^2_{\ms}\overline{\sigma}^2_{\ms}}\right)\right]=\frac{2N_c\Delta^4}{(4\pi)^2}\left[\frac{1}{\epsilon}+\frac{3}{2}+\ln\left(\frac{\Lambda^2}{\Delta^2}\right)\right]\;,
\eqa \\
One can define the sacle $\Lambda$ independent parameter $\Delta=\frac{g_{\ms}\overline{\sigma}_{\ms}}{2}$ using the Eq.(\ref{params6}) and Eq. (\ref{params7}).It is instructive to write the Eq.(\ref{omegams1}) in terms of the sacle independent $\Delta$ as \\ 
\bqa
\nonumber
U(\Delta)&=&2\frac{m_{\ms}^2(\Lambda)}{g^2_{\ms}(\Lambda)}\Delta^2-2\frac{c_{\ms}(\Lambda)}{g^2_{\ms}(\Lambda)}\Delta^2+4\left(\frac{\lambda_{1\ms}(\Lambda)}{g^4_{\ms}(\Lambda)}+\frac{\lambda_{2\ms}(\Lambda)}{2g^4_{\ms}(\Lambda)}\right)\Delta^4-2\frac{h_{\ms}(\Lambda)}{g_{\ms}(\Lambda)}\Delta\; \\
&=&2\left(\frac{m^2_0}{g^2_0}-\frac{c_0}{g^2_0}\right)\Delta^2+4\left(\frac{\lambda_{10}}{g^4_0}+\frac{\lambda_{20}}{2g^4_0}\right)\Delta^4-2\frac{h_0}{g_0}\Delta
\eqa

\bqa
\Omega_{vac}(\Delta)&=&2\left(\frac{m^2_0}{g^2_0}-\frac{c_0}{g^2_0}\right)\Delta^2+4\left(\frac{\lambda_{10}}{g^4_0}+\frac{\lambda_{20}}{2g^4_0}\right)\Delta^4-2\frac{h_0}{g_0}+\frac{2N_c\Delta^4}{(4\pi)^2}\left[\frac{3}{2}+\ln\left(\frac{\Lambda^2}{\Delta^2}\right)\right]\;
\eqa
When the couplings and mass parameter are expressed in terms of the physical meson masses, pion decay constant and Yukawa coupling, one can write 
\bqa
\label{rqmeff}
\nonumber
\Omega_{vac}(\Delta)
&=&\frac{(3m^2_\pi-m^2_\sigma)f^2_{\pi}}{4}\left\lbrace 1-\frac{N_cg^2}{(4\pi)^2}\left(\mathcal{C}(m^2_\pi)+m^2_\pi\mathcal{C}^\prime(m^2_\pi)\right)\right\rbrace\frac{\Delta^2}{m^2_q}+\frac{N_cg^2f^2_\pi}{2(4\pi)^2}\left\lbrace\frac{3m^2_\pi\mathcal{C}(m^2_\pi)-(m^2_\sigma-4m^2_q)\mathcal{C}(m^2_\sigma)}{2}-2m^2_q\right\rbrace\frac{\Delta^2}{m^2_q}\;\\ \nonumber
&&+\frac{(m^2_\sigma-m^2_\pi)f^2_\pi}{8}\left\lbrace1-\frac{N_cg^2}{(4\pi)^2}\left(\mathcal{C}(m^2_\pi)+m^2_\pi\mathcal{C}^\prime(m^2_\pi)\right)\right\rbrace\frac{\Delta^4}{m^4_q}+\frac{N_cg^2f^2_\pi}{(4\pi)^2}\left[(m^2_\sigma-4m^2_q)\mathcal{C}(m^2_\sigma)-m^2_\pi\mathcal{C}(m^2_\pi)\over 8\right]\frac{\Delta^4}{m^4_q}\;\\
&&+\frac{2N_c\Delta^4}{(4\pi)^2}\left\lbrace\frac{3}{2}-\ln\left(\frac{\Delta^2}{m^2_q}\right)\right\rbrace-m^2_\pi f^2_\pi\left\lbrace 1-\frac{N_cg^2}{(4\pi)^2}m^2_\pi\mathcal{C}^\prime(m^2_\pi)\right\rbrace\frac{\Delta}{m_q}\;
\eqa
\end{widetext}
Here we point out that when we get the final expression of the RQM model vacuum effective potential to one-loop order after renormalization and consistent parameter fixing, the $m_\eta$ and $m_{a_0}$ dependent correction factors cancel out.
We have checked that the expression (\ref{rqmeff}) turns out to be equivalent to the expression of the vacuum effective potential calculated in the Ref. \cite{Adhiand2} (Eq.(7) with $q=0$) and also given in the Eq.(38) of the Ref. \cite{asmuAnd}. 

\begin{widetext}
\bqa
\nonumber
\label{rqmomega}
\Omega_{RQM}(\Delta,T,\mu)&=&\frac{(3m^2_\pi-m^2_\sigma)f^2_{\pi}}{4}\left\lbrace 1-\frac{N_cg^2}{(4\pi)^2}\left(\mathcal{C}(m^2_\pi)+m^2_\pi\mathcal{C}^\prime(m^2_\pi)\right)\right\rbrace\frac{\Delta^2}{m^2_q}\; \\ \nonumber
&&+\frac{N_cg^2f^2_\pi}{2(4\pi)^2}\left\lbrace\frac{3m^2_\pi\mathcal{C}(m^2_\pi)-(m^2_\sigma-4m^2_q)\mathcal{C}(m^2_\sigma)}{2}-2m^2_q\right\rbrace\frac{\Delta^2}{m^2_q}\; \\ \nonumber 
&&+\frac{(m^2_\sigma-m^2_\pi)f^2_\pi}{8}\left\lbrace1-\frac{N_cg^2}{(4\pi)^2}\left(\mathcal{C}(m^2_\pi)+m^2_\pi\mathcal{C}^\prime(m^2_\pi)\right)\right\rbrace\frac{\Delta^4}{m^4_q}\; \\ \nonumber
&&+\frac{N_cg^2f^2_\pi}{(4\pi)^2}\left[(m^2_\sigma-4m^2_q)\mathcal{C}(m^2_\sigma)-m^2_\pi\mathcal{C}(m^2_\pi)\over 8\right]\frac{\Delta^4}{m^4_q}+\frac{2N_c\Delta^4}{(4\pi)^2}\left\lbrace\frac{3}{2}-\ln\left(\frac{\Delta^2}{m^2_q}\right)\right\rbrace\; \\ \nonumber
&&-m^2_\pi f^2_\pi\left\lbrace 1-\frac{N_cg^2}{(4\pi)^2}m^2_\pi\mathcal{C}^\prime(m^2_\pi)\right\rbrace\frac{\Delta}{m_q}-4N_cT\int \frac{d^3p}{(2\pi)^3}\left\lbrace \ln\left[1+e^{-E_{q}^{+}/T)}\right]+\ln\left[1+e^{-E_{q}^{-}/T)}\right]\right\rbrace\; \\ 
\eqa
\end{widetext}

One gets the chiral condensate or the parameter $\Delta$ in the RQM model by searching the global 
minima of the grand potential in the Eq.(\ref{rqmomega}) for a given value of 
temperature T and chemical potential $\mu$

\begin{equation}
\frac{\partial \Omega_{RQM}(\Delta,T,\mu)}{\partial
      \Delta} =0.
\label{EoMMF3}
\end {equation}

In our calculations we have used the $m_\pi=138.0$ MeV,$m_{a_0}=984.7$ MeV and $m_\eta=547.0$ MeV,The Yukawa coupling $g=6.5$ and pion decay constant $f_\pi=93.0$ MeV.The constituent quark mass in the vacuum $m_q=\frac{gf_\pi}{2} = 302.25 $ MeV.

\begin{table}[!htbp]
    \caption{Critical temperature for $m_\sigma=$400, 500, 600, 616 and 700 MeV.}
    \label{tab:table2}
    \begin{tabular}{p{2cm} p{2cm} p{2cm} p{2cm} }
      \toprule 
      $m_{\sigma}(\text{MeV})$&$T_c\text{(QM)}$ & $T_c\text{(QMVT)}$ & $T_c\text{(RQM)}$\\
      \hline 
      \hline
      $400$ & $113.3$ & $143.6$ & $131.8$\\
      $500$ & $130.2$ & $157.3$ & $145.6$\\
      $600$ & $147.8$ & $173.1$ & $169.3$\\
      $616$ & $150.5$ & $175.6$ & $175.6$\\
      $700$ & $166.1$ & $189.8$ & $203.6$\\
      \hline 
    \end{tabular}
\end{table}

\section{Results and Discussion}

\begin{figure*}[!htbp]
\subfigure[\ Effective potential for $m_\sigma=500 \ \text{MeV}$.]{
\label{fig:mini:fig6:a} 
\begin{minipage}[b]{0.32\textwidth}
\centering \includegraphics[width=\linewidth]{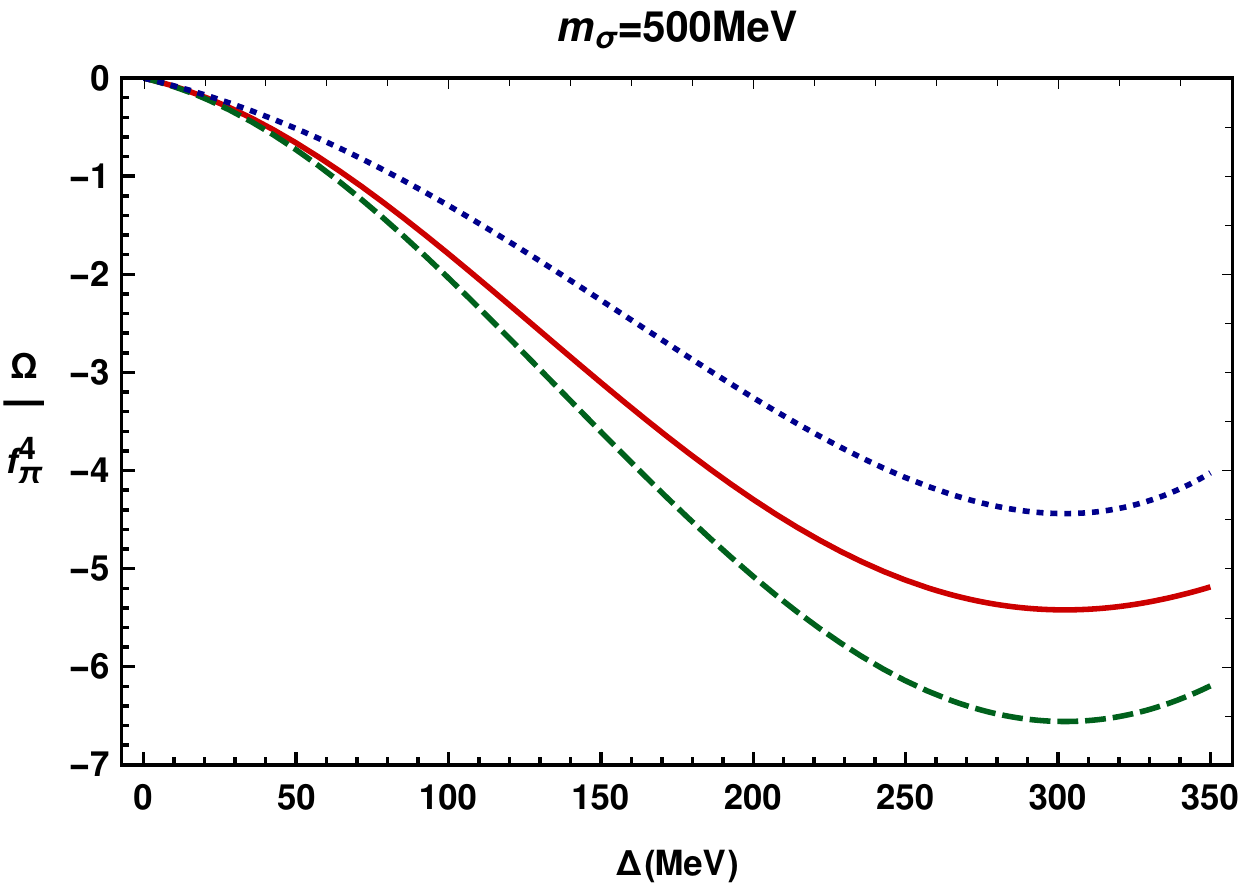}
\end{minipage}}%
\hfill
\subfigure[\ Effective potential for $m_\sigma=616 \ \text{MeV}$.]{
\label{fig:mini:fig6:b} 
\begin{minipage}[b]{0.32\textwidth}
\centering \includegraphics[width=\linewidth]{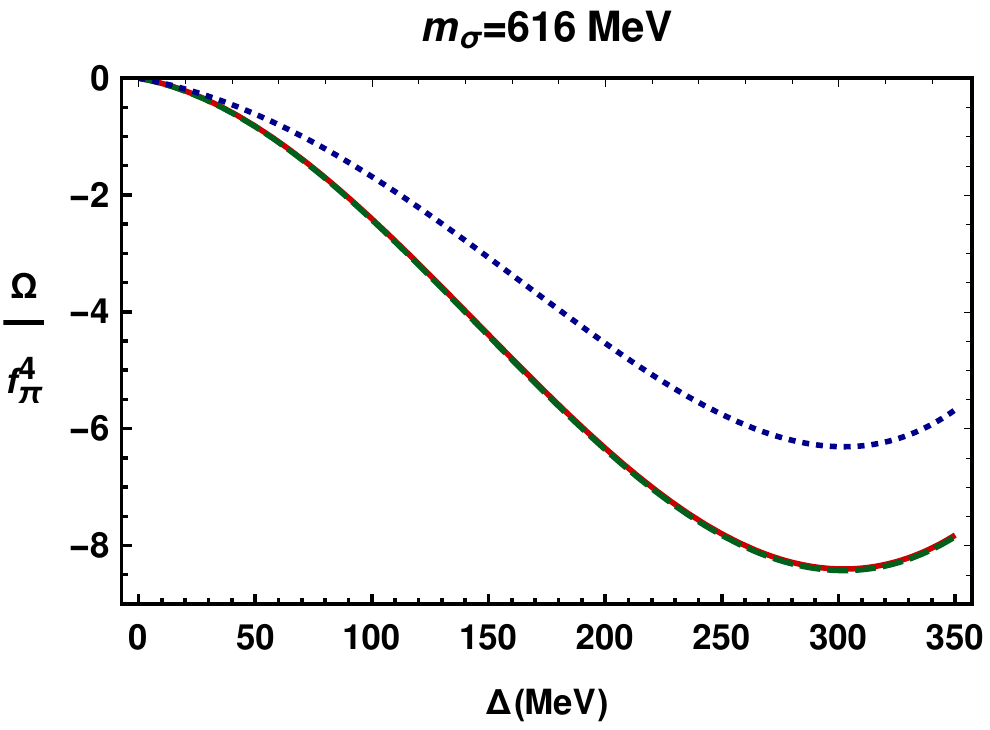}
\end{minipage}}
\hfill
\subfigure[\ Effective potential for $m_\sigma=700 \ \text{MeV}$.]{
\label{fig:mini:fig6:c} 
\begin{minipage}[b]{0.32\textwidth}
\centering \includegraphics[width=\linewidth]{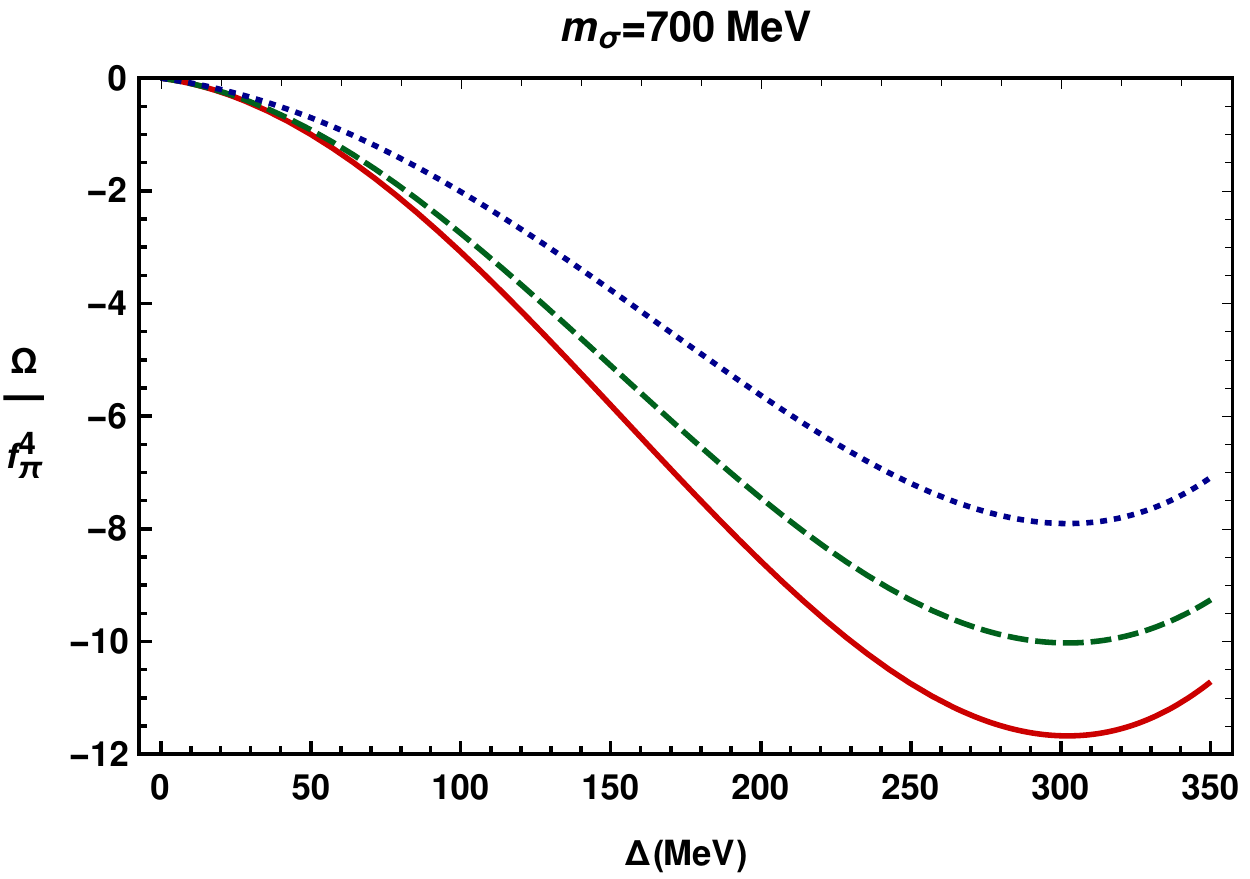}
\end{minipage}}
\caption{Blue dotted line,solid red line and green dashed line respectively depict the QM,RQM and QMVT model result.}
\label{fig:mini:fig6} 
\end{figure*} 

\begin{figure*}[!htbp]
\subfigure[\ Chiral order parameter  $m_\sigma=500 \ \text{MeV}$.]{
\label{fig:mini:fig7:a} 
\begin{minipage}[b]{0.325\textwidth}
\centering \includegraphics[width=\linewidth]{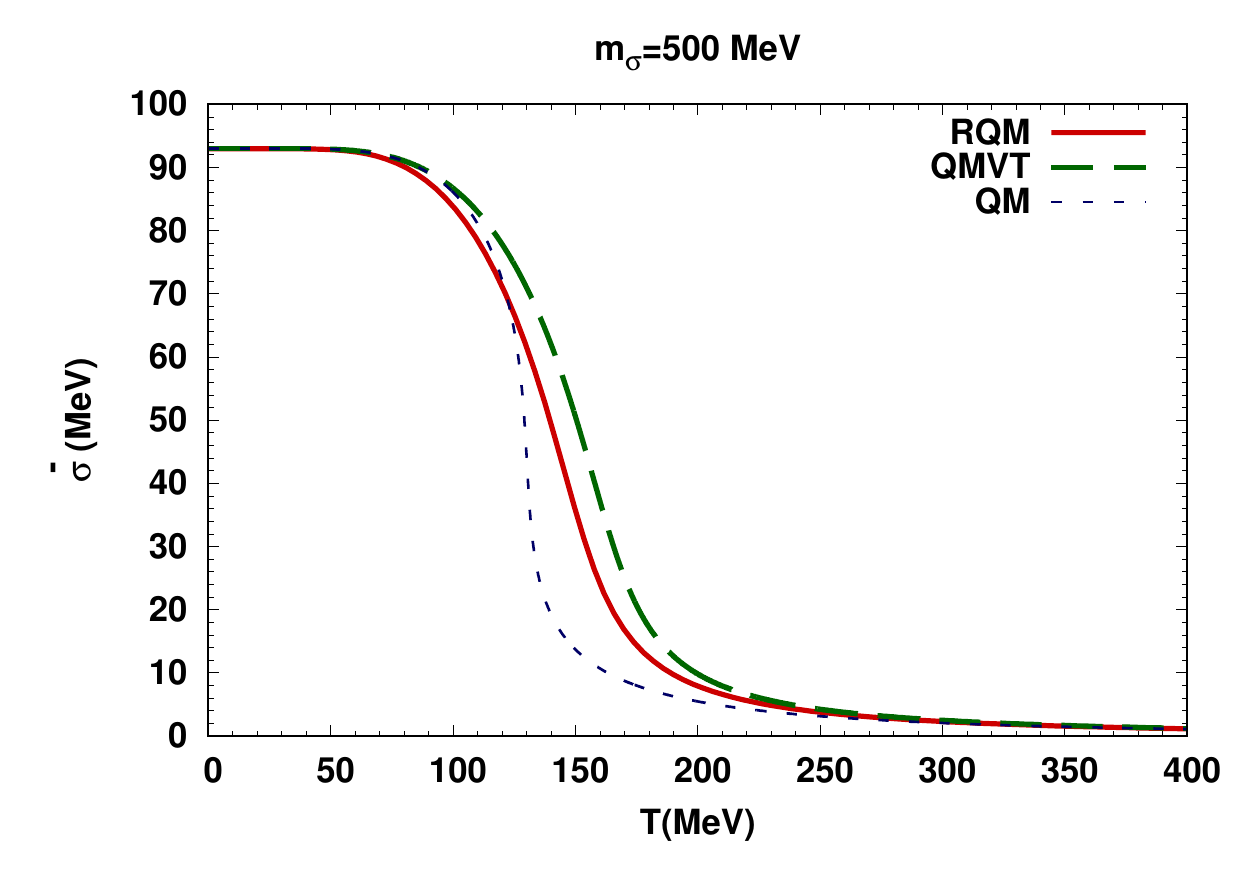}
\end{minipage}}%
\hfill
\subfigure[\ Chiral order parameter $m_\sigma=616 \ \text{MeV}$.]{
\label{fig:mini:fig7:b} 
\begin{minipage}[b]{0.325\textwidth}
\centering \includegraphics[width=\linewidth]{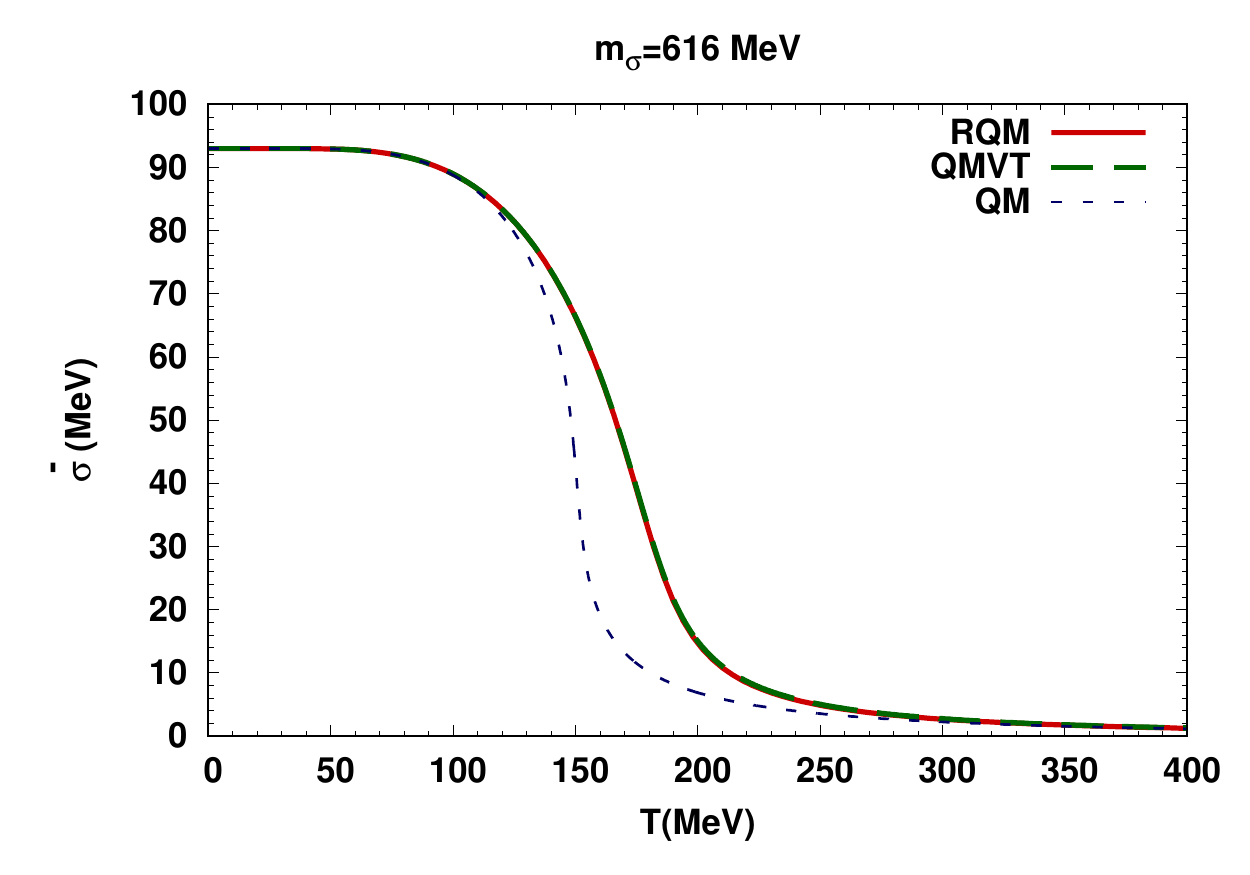}
\end{minipage}}
\hfill
\subfigure[\ Chiral order parameter $m_\sigma=700 \ \text{MeV}$.]{
\label{fig:mini:fig7:c} 
\begin{minipage}[b]{0.325\textwidth}
\centering \includegraphics[width=\linewidth]{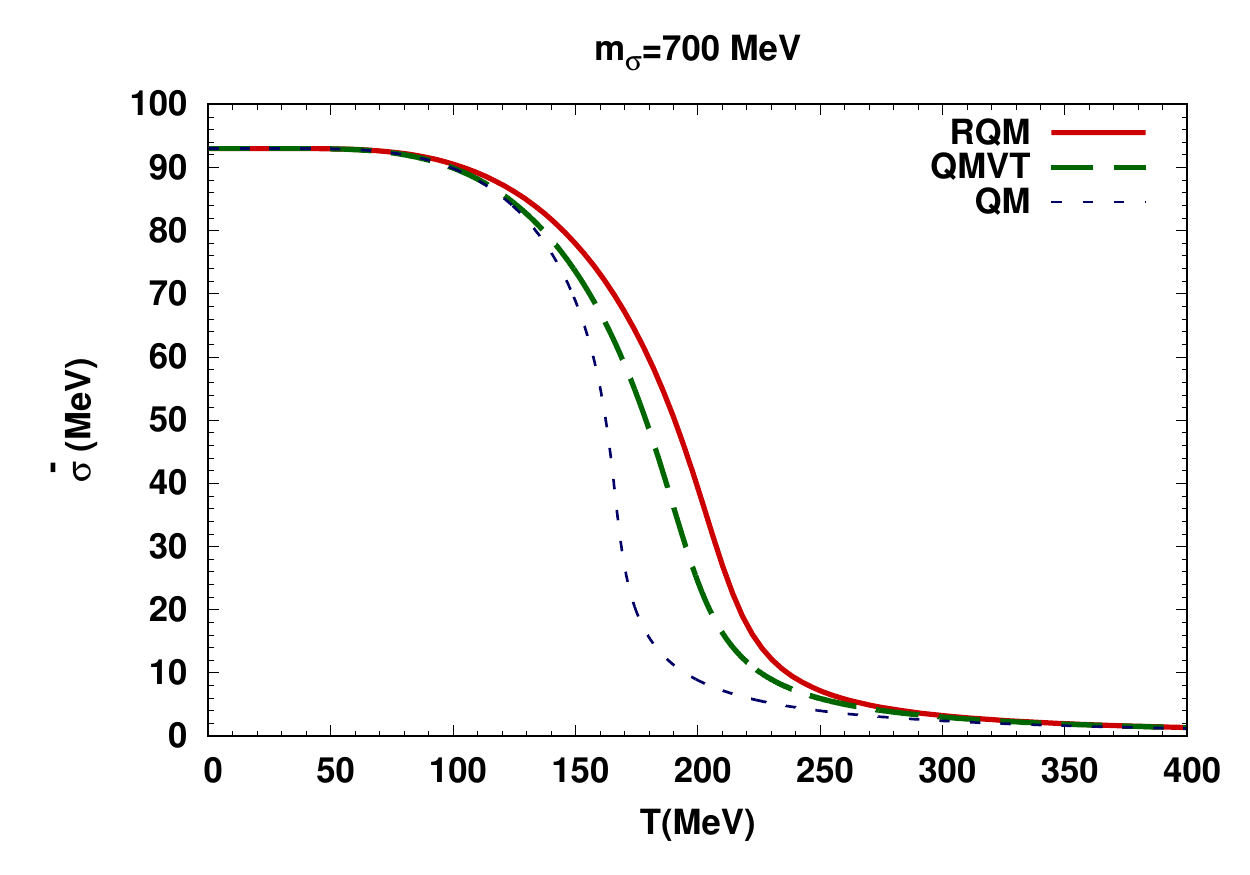}
\end{minipage}}
\caption{Blue dotted line,solid red line and green dashed line present the respective temperature variation in the QM,RQM and QMVT model.}
\label{fig:mini:fig7} 
\end{figure*} 

\begin{figure}[!htbp]
\label{fig:mini:fig8} 
\centering \includegraphics[width=\linewidth]{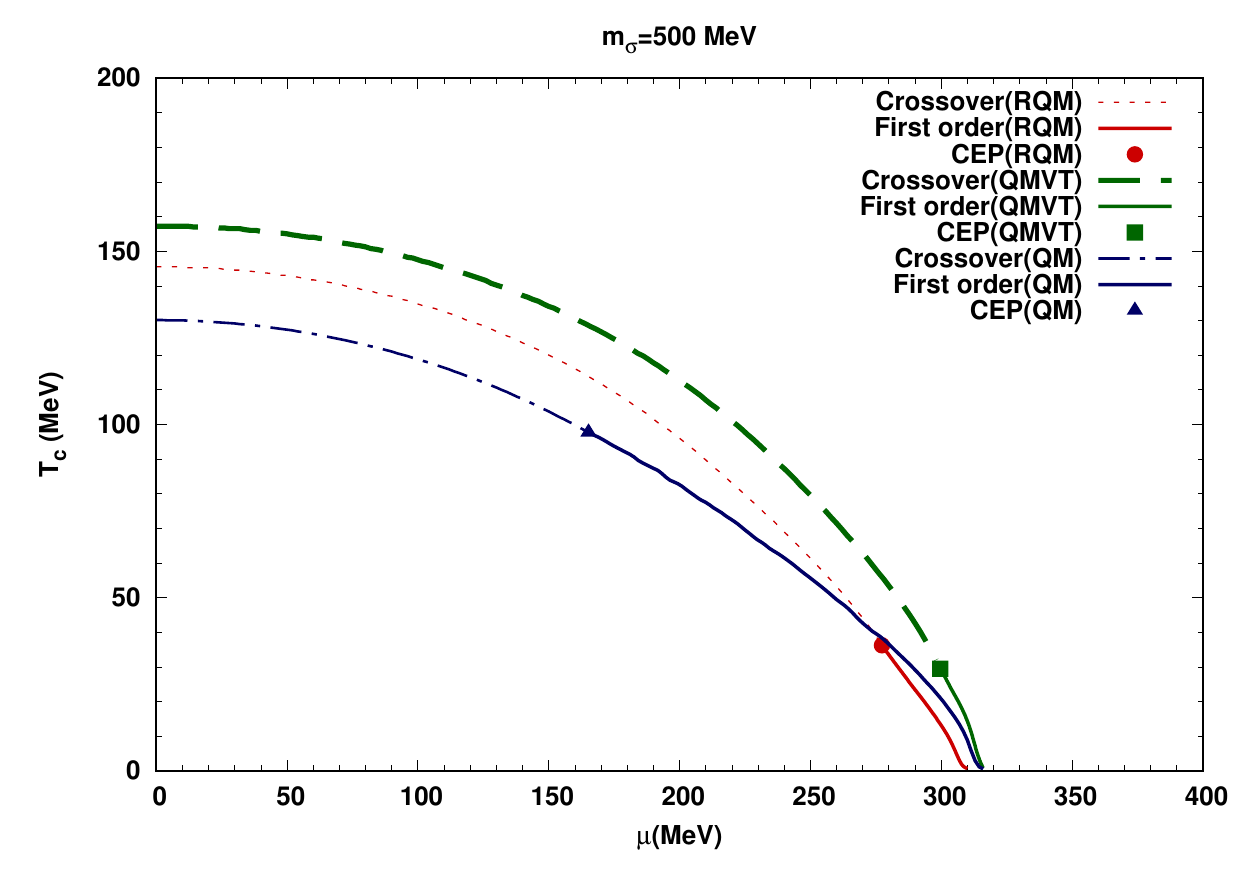}%
\caption{Phase diagram for the $m_\sigma=500$ MeV in the QM, RQM and QMVT model.}
\label{fig:mini:fig8} 
\end{figure}

\begin{figure}[!htbp]
\centering \includegraphics[width=\linewidth]{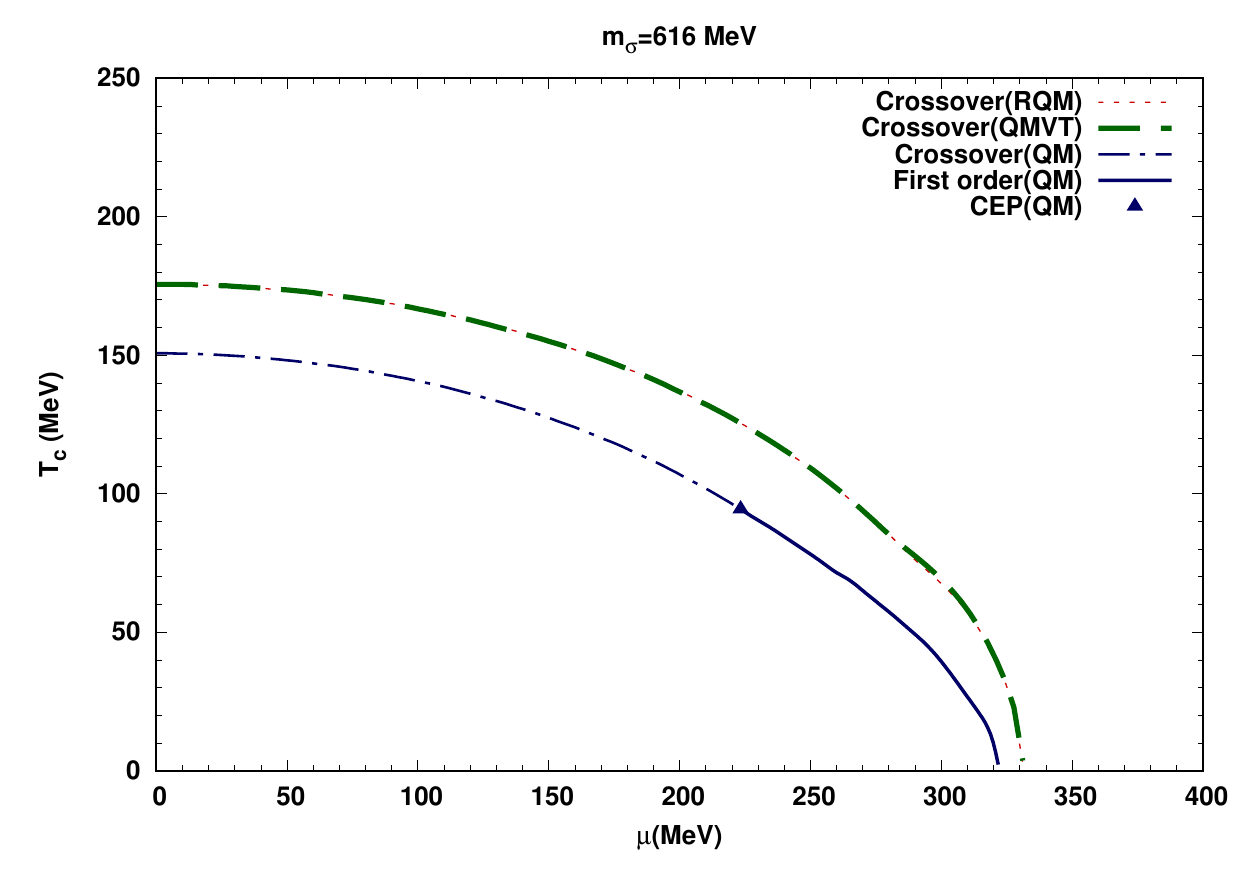}%
\caption{Phase diagram for the $m_\sigma=616$ MeV in the QM, RQM and QMVT model.}
\label{fig:mini:fig9} 
\end{figure}

\begin{figure}[!htbp]
\centering \includegraphics[width=\linewidth]{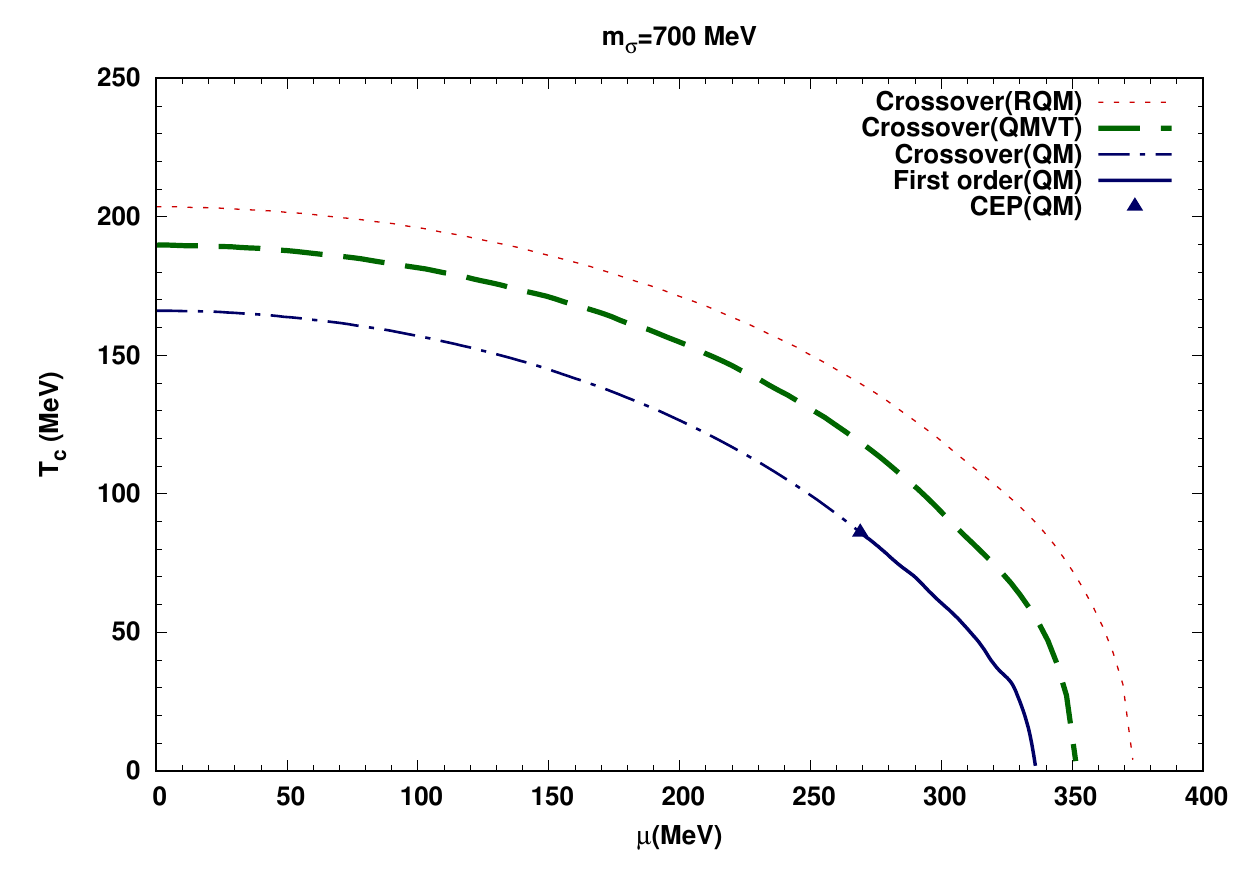}%
\caption{Phase diagram for the $m_\sigma=700$ MeV in the QM, RQM and QMVT model.}
\label{fig:mini:fig10} 
\end{figure}

\begin{figure}[!htbp]
\centering \includegraphics[width=\linewidth]{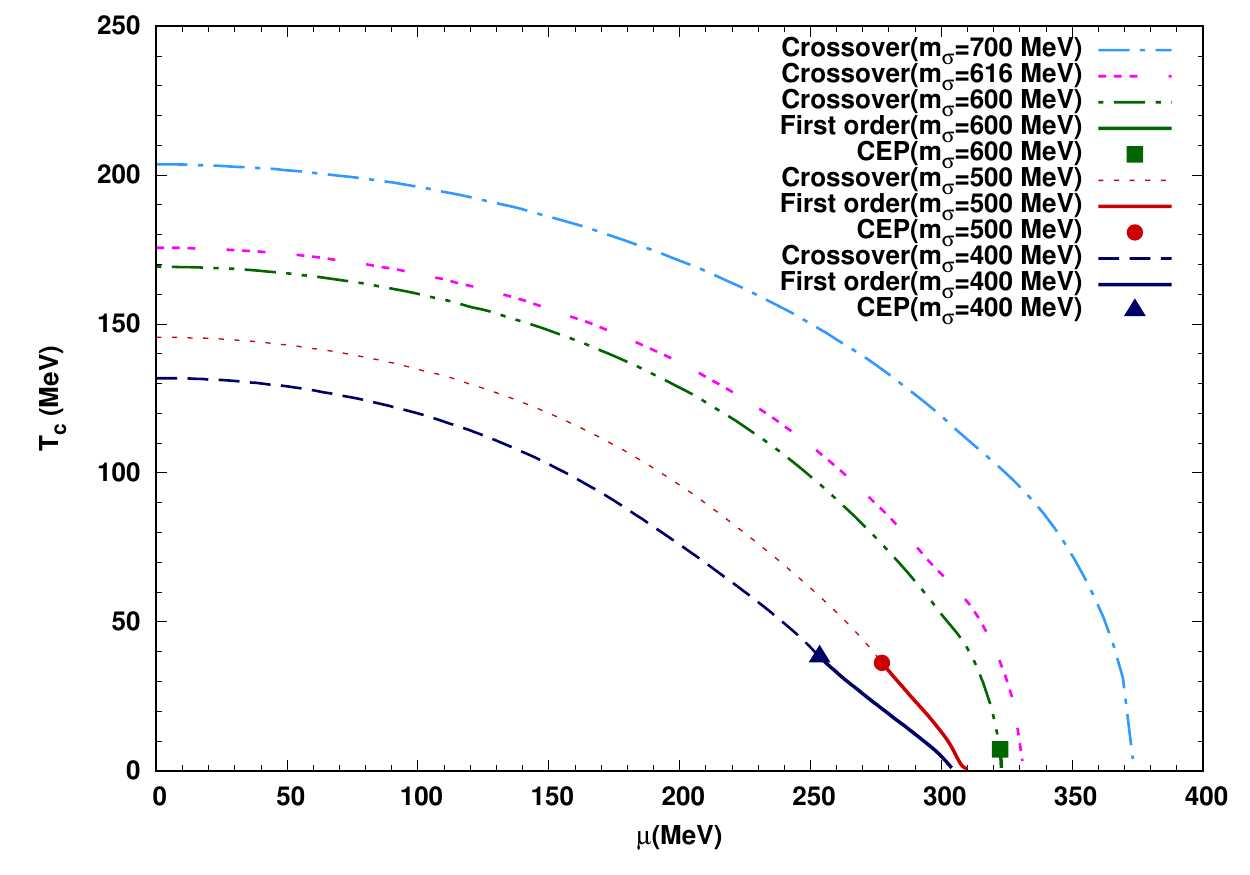}%
\caption{Phase diagram for the sigma masses of 400, 500, 600, 616 and 700 MeV in the RQM model.}
\label{fig:mini:fig11} 
\end{figure}

We have potted the normalized effective potential in the vacuum at $\mu=0$ and $T=0$ with respect to the constituent quark mass scale independent parameter $\Delta$ in Fig.(\ref{fig:mini:fig6}) for the different model scenarios of the parameter fixing and for the different sigma meson masses.The RQM model result is depicted by the solid line in red,dotted line in blue shows the QM model plot and dashed green line is plotting the QMVT model result.The effective potential plots corresponding to the $m_\sigma=$ 500 MeV, 616 MeV and 700 MeV are presented respectively in the Fig.(\ref{fig:mini:fig6:a}), (\ref{fig:mini:fig6:b}) and (\ref{fig:mini:fig6:c}).The minimum of the effective potential in all the Figs. for every model scenario, occurs at $\Delta=302.25=m_q$ MeV and its value is highest (i.e. it is most shallow) in the case of no-sea approximation of the QM model for all the $m_\sigma$ values.It is evident from the Fig.(\ref{fig:mini:fig6:a}) that for the $m_\sigma=$ 500 MeV,the effective potential is deepest for the QMVT model parameter fixing while the on-sell parameterization in the RQM model gives a shallower effective potential in comparison.Interestingly as we increase the $m_\sigma$ value,we notice that the RQM model effective potential becomes deeper while the QMVT model effective potential shows an upward trend and finally for the $m_\sigma=616 \ \text{MeV}$,the effective potential plots for both the model scenarios coincide with each other as shown in the Fig.(\ref{fig:mini:fig6:b}).Increasing the $m_\sigma$ value beyond the 616 MeV,the effective potential plot becomes deepest for the RQM model.The plots of Fig.(\ref{fig:mini:fig6:c}) for the $m_\sigma=700 \text{MeV}$,show that the effective potential becomes deepest for the RQM model and in reversal of the trend seen in the Fig.(\ref{fig:mini:fig6:a}),here the effective potential of the QMVT model  is shallower than that of the RQM model.

 We have potted the temperature variation of the quark condensate $\overline \sigma$ (which is obtianed from the $\Delta$ as the Yukawa coupling $g$ remains the same after renormalization) at $\mu=0$ in Fig.(\ref{fig:mini:fig7}) for three values of $m_\sigma=$ 500,616 and 700 MeV.Here also the solid red line,the blue dotted line and the dashed green line respectively plot the RQM model,the QM model and the QMVT model results.In general,the chiral transition becomes smoother due to the quark one-loop vacuum correction.For the $m_\sigma=$ 500 MeV case in the Fig.(\ref{fig:mini:fig7:a}),the sharpest QM model quark condensate temperature variation  becomes more smooth for the on-shell parameterization of the RQM model while the most smooth variation of the condensate is seen in the QMVT model plot.Sharpest chiral crossover transition occurs early at a pseudo-critical temperature of $T_c=130.2$ MeV in the QM model and a smoother chiral crossover is witnessed for the RQM model at $T_c=145.6$ MeV while a most delayed and smooth chiral crossover occurs at $T_c=157.3$ MeV in the QMVT model.For the $m_\sigma=$ 616 MeV case in the Fig.(\ref{fig:mini:fig7:b}),the RQM model result exactly coincides with the QMVT model result for the temperature variation of the quark condensate and the chiral crossover transition occurs at the same temperature of $T_c=175.6$ MeV.We notice that this behaviour follows from the complete coincidence of the vacuum effective potential plot of the RQM model with that of the QMVT model in Fig.(\ref{fig:mini:fig6:b}) for the $m_\sigma=$ 616 MeV case.For the $m_\sigma=$ 700 MeV in the Fig.(\ref{fig:mini:fig7:c}),the most smooth temperature variation of the quark condensate occurs in the RQM model with a very delayed crossover transition at  $T_c=203.6$ MeV while 
 we notice that a less smooth chiral crossover transition occurs earlier at $T_c=189.8$ MeV in the QMVT model.We point out that when we compare the quark condensate temperature variation of the RQM model with that of the QMVT model,we find that for the $m_\sigma=$ 700 MeV the trend  becomes opposite of what we observe for the $m_\sigma=$ 500 MeV case in the Fig.(\ref{fig:mini:fig7:a}) where the RQM model condensate variation is less smooth and sharper than the QMVT model condensate variation.Here it is relevant to remind that the vacuum effective potential depth for the RQM model when compared with that of the QMVT model,shows the similar role reversal when the $m_\sigma=$ 500 MeV plots are contrasted with the $m_\sigma=$ 700 MeV plots.
 
We have drawn the $\mu-T$ plane phase diagram for the $m_\sigma=500$ MeV in Fig.(\ref{fig:mini:fig8}) with labelled line types.The QM model critical end point (CEP) location at $\mu=$165.2 MeV, $T=$97.7 MeV shifts to a far right position in the $\mu-T$ plane at $\mu=$299.6 MeV, $T=$29.48 MeV due to the quark one loop vacuum correction in the QMVT model setting.Earlier studies \cite{guptiw,schafwag12,chatmoh1,vkkr12,chatmoh2} reporting similar results have concluded that incorporating the fermionic vacuum fluctuation in the QM model,leads to a robust and significant change in the location of CEP.Here we point out that,the exact on-shell renormalization of the quark one-loop vacuum fluctuation for the parameter fixing in the RQM model,gives a phase diagram in which the CEP location $\mu=$277.3 MeV, $T=$36.2 MeV is at lower chemical potential and higher temperature i.e. CEP shifts higher up when compared to the position of the CEP in the QMVT model.Furthermore,the RQM model phase diagram for the $m_\sigma=500$ MeV,stands in the immediate proximity of the QM model phase diagram.

Here,we emphasize the interesting observation that the RQM model phase diagram,depicting a crossover transition line in the whole of the $\mu-T$ plane,coincides exactly with the QMVT model phase diagram in the Fig.(\ref{fig:mini:fig9}).This overlap follows from the exact coincidence of the vacuum effective potential plots in the Fig.(\ref{fig:mini:fig6:b}) for both the models RQM and QMVT when the $m_\sigma$=616 MeV.The first order line of the QM model plot in the Fig.(\ref{fig:mini:fig9}) is ending in the critical end point (CEP) at $\mu=$223.3 MeV, $T=$94.45 MeV.In comparison to the RQM model phase diagram,the QMVT model phase diagram,in Fig.(\ref{fig:mini:fig10}) for the $m_\sigma=700$ MeV,stands closer to the QM model phase diagram.It shows the usual trend reversal when compared to the $m_\sigma=500$ MeV case plots.Again this behaviour follows from the trend reversal that we observe in the behaviour of 
the vacuum effective potential in the Fig.(\ref{fig:mini:fig6:c}).We get a crossover line in
the whole $\mu-T$ plane for both the models the RQM as well as QMVT in the Fig.(\ref{fig:mini:fig10}) while the first order line of the QM model phase diagram  is terminating at the critical end point (CEP) at $\mu=$223.3 MeV, $T=$94.45 MeV.

In order to see the effect of the sigma meson mass,Fig.(\ref{fig:mini:fig11}) plots the five phase diagrams 
for the $m_\sigma$= 400 MeV, 500 MeV, 600 MeV, 616 MeV and 700 MeV in the RQM model.The CEP, $\mu$=253.5 MeV, $T$=38.2 MeV for the $m_\sigma$= 400 MeV,moves rightwards to $\mu$=277.3 MeV, $T$=36.2 MeV for the $m_\sigma$= 500 MeV.When the $m_\sigma$= 600 MeV,the CEP shifts to the extreme right bottom of the phase diagram at $\mu$=322.5 MeV, $T$=7.2 MeV.The phase transition line becomes a crossover in the whole of the $\mu-T$ plane for the $m_\sigma$ = 616 MeV.We have also shown the crossover phase transition line for the $m_\sigma$ = 700 MeV.

\section{Summary and Conclusion}

We have applied the on-shell parameter fixing prescription,to the quark-meson (QM) model in which the two flavor of quarks are coupled to the  eight mesons of the $ SU(2)_{L} \times SU(2)_{R} $ linear sigma model,and then calculated the one-loop effective potential for the renormalized quark-meson (RQM) model whose six running parameters $\lambda_{1}$, $\lambda_{2}$, $c$, $m^{2}$, $h$, $g$ are determined by relating the $\overline{\text{MS}}$, on-shell schemes and the experimental values of the quark, meson masses and pion decay constant.After including the one quark-loop vacuum correction in the QM model,the effective potential has been calculated also when the curvature meson masses are used for fixing the model parameters and this  model setting has been termed as the quark-meson model with the vacuum term (QMVT).We have computed and compared the effective potentials,the order parameter temperature variations and the phase diagrams for the QM, RQM and QMVT model settings.

The differences and similarities for the vacuum effective potential plots in the RQM model and the QMVT model,depend on the sigma meson mass.When we plot the normalized effective potential with respect to the constituent quark mass parameter $\Delta$,its depth is highest for the QMVT model if the $m_\sigma=$500 MeV and the effective potential is less deep and least in depth respectively for the RQM model and QM model.For the $m_\sigma=$616 MeV,the QMVT model and the RQM model effective potentials become exactly identical to each other.For the higher $m_\sigma=$700 MeV,the effective potential of the RQM model becomes most deep and the interesting trend reversal is noticed when one contrasts it with the variation of the effective potential for the $m_\sigma=$500 MeV case.Comparing the $\mu=0$ temperature variations of the order parameter,for the QM, RQM and QMVT model settings,we find exactly similar differences and similarities on the sigma meson mass as observed in the nature of the corresponding normalized effective potential.

It is well reported in the earlier research literature \cite{guptiw,schafwag12,chatmoh1,vkkr12,chatmoh2}, that incorporating the quark one-loop vacuum correction in the QMVT model setting,gives rise to a phase diagram in which the CEP shifts towards the right side of the $\mu-T$ plane,to quite a higher value of the chemical potential and a lower value of the temperature when one compares it with the location of the CEP in the QM model phase diagram.We have found that when the $m_\sigma=$500 MeV,the shift in the position of the CEP observed in the RQM model,is smaller than what is observed in the QMVT model and the RQM model phase boundary stands closer to the QM model phase diagram.Furthermore,driven directly by the nature of the effective potential variation,the phase boundaries depicting the crossover transition lines,for both the models RQM and QMVT,completely overlap with each other when the $m_\sigma=$616 MeV.For the higher $m_\sigma=$700 MeV,the crossover line of the QMVT model phase diagram,comes closer to the QM model phase boundary.This trend is opposite of what we see for the $m_\sigma=$500 MeV case when the RQM model phase boundary stands closer to the QM model phase diagram.Again the above behaviour is caused by the corresponding reversal in the variation of the normalized vacuum effective potential.

\section*{Acknowledgments}
Computational support of the computing facility which has been developed by the Nuclear Particle Physics group of the Department of Physics, University of Allahabad (UOA) under the Center
 of Advanced Studies (CAS) funding of UGC, India, is acknowledged.Department of Science and Technology,Government of India, DST-PURSE programme Phase 2/43(C), financial support to the science faculty of the
 UOA is also acknowledged.

\appendix
\section{THE QMVT PARAMETER FIXING}
\label{appenA}
In the QMVT model, the meson masses at $T=0$ and $\mu=0$, get contribution from the second derivative of the pure mesonic potential  $U$ and the second derivative of $\Omega^{vac}_{q\bar q}$. It is written as
\bqa
\label{m2meson}
m^2_{\alpha,ab}&=\left.\frac{\partial^2 \Omega^\Lambda(\overline{\sigma})}{\partial \xi_{\alpha,a}\partial \xi_{\alpha,b}}\right|_{min}=&(m^m_{\alpha,ab})^2+(\delta m^v_{\alpha,ab})^2
\eqa
where $\alpha=s,p$; ``$s$'' stands for the scalar and ``$p$'' stands for the pseudoscalar mesons and $a,b=0,1,2,3$.
$m^2_{s,00}\equiv m^2_\sigma$; $m^2_{s,11}=m^2_{s,22}=m^2_{s,33}\equiv m^2_{a_0}$ and $m^2_{p,00}\equiv m^2_\eta$; $m^2_{p,11}=m^2_{p,22}=m^2_{p,33}\equiv m^2_{\pi}$.The $(m^m_{\alpha,ab})^2$ and $(\delta m^v_{\alpha,ab})^2$  are defined in the similar fashion, superscript ``$m$'' stands for the pure mesonic contribution and ``$v$'' stands for quark/qntiquark vaccum contribution.Here$\left.\right|_{min}$ represents the global minimum of the grand potential.
\begin{table}[!htbp]
  \begin{center}
    \caption{Expressions of the curvature masses $(m^m_{\alpha,ab})^2$ are calculated from the second derivative of the pure mesonic potential as has been evaluated in Ref.\cite{Roder}.}
    \label{tab:table3}
    \begin{tabular}{p{2cm} p{2cm} p{3.7cm}}
      \toprule 
      $(m^m_{\alpha,ab})^2$&& Meson mass found from the pure mesonic potential  \\
      \hline 
      \hline
      $(m^m_{s,00})^2$&$(m^m_\sigma)^2$& $m^2-c+3\left(\lambda_1+\frac{\lambda_2}{2}\right)\overline{\sigma}^2$\\
      $(m^m_{s,11})^2$&$(m^m_{a_0})^2$&$m^2+c+\left(\lambda_1+\frac{3\lambda_2}{2}\right)\overline{\sigma}^2$\\
      $(m^m_{p,00})^2$&$(m^m_\eta)^2$&$m^2+c+\left(\lambda_1+\frac{\lambda_2}{2}\right)\overline{\sigma}^2$\\
      $(m^m_{p,11})^2$&$(m^m_\pi)^2$& $m^2-c+\left(\lambda_1+\frac{\lambda_2}{2}\right)\overline{\sigma}^2$\\
      \hline 
    \end{tabular}
  \end{center}
\end{table}
\bqa
\label{mvac}
\nonumber
(\delta m^v_{\alpha,ab})^2&=&\left.\frac{\partial^2 \Omega^{vac}_{q\bar q}}{\partial \xi_{\alpha,a} \partial \xi_{\alpha,b}}\right|_{min}\;,\\ 
\nonumber
&=&\sum_{q=u,d} \frac{2N_c}{(4\pi)^2}\left[\left\lbrace m^2_{q,\alpha a}m^2_{q,\alpha b}+m^2_qm^2_{q,\alpha ab}\right\rbrace\right. \\
&&\left.\left\lbrace 1+\ln\left(\frac{\Lambda^2}{m^2_q}\right)\right\rbrace-m^2_{q,\alpha a}m^2_{q,\alpha b}\right] \;.
\eqa
where $m^2_{q,\alpha a}=\frac{\partial m^2_q}{\partial \xi_{\alpha,a}}$ and $m^2_{q,\alpha ab}=\frac{\partial m^2_{q,\alpha a}}{\partial \xi_{\alpha,b}}$.
\begin{table}[!htbp]
  \begin{center}
    \caption{The last two columns present the first and second derivative of the squared quark mass summed over two quark flavor.}
    \label{tab:table4}
    \begin{tabular}{p{1cm} p{1cm} p{1cm} p{2.5cm} p{2.5cm}}
      \toprule 
      $s/p$&$a$&$b$  & $m^2_{q,\alpha a}m^2_{q,\alpha b}/g^4$ & $m^2_{q,\alpha ab}/g^2$\\
      \hline 
      \hline
      $s$& $0$ &$0$ & $\frac{1}{2}{\overline{\sigma}}^2$ & 1\\
      $s$& $1$ &$1$ & $\frac{1}{2}{\overline{\sigma}}^2$ & 1\\
      $p$& $0$ &$0$ & 0 & 1\\
      $p$& $1$ &$1$ & 0 & 1\\
      \hline 
    \end{tabular}
  \end{center}
\end{table}
Using the Table-\ref{tab:table4} in the Eq.(\ref{mvac}) we get vacuum contributions of meson masses as,
\bqa
\label{vacmass1}
(\delta m^v_\sigma)^2&\equiv&(\delta m^v_{s,00})^2=\dfrac{N_cg^4\overline{\sigma}^2}{2(4\pi)^2}\left[1+3\ln\left(\dfrac{\Lambda^2}{m^2_q}\right)\right]\;, \
\eqa
\bqa
(\delta m^v_{a_0})^2&\equiv&(\delta m^v_{s,11})^2=\dfrac{N_cg^4\overline{\sigma}^2}{2(4\pi)^2}\left[1+3\ln\left(\dfrac{\Lambda^2}{m^2_q}\right)\right]\;, \
\eqa
\bqa
(\delta m^v_\eta)^2&\equiv&(\delta m^v_{p,00})^2=\dfrac{N_cg^4\overline{\sigma}^2}{2(4\pi)^2}\left[1+\ln\left(\dfrac{\Lambda^2}{m^2_q}\right)\right]\;, \
\eqa
\bqa
(\delta m^v_\pi)^2&\equiv&(\delta m^v_{p,11})^2=\dfrac{N_cg^4\overline{\sigma}^2}{2(4\pi)^2}\left[1+\ln\left(\dfrac{\Lambda^2}{m^2_q}\right)\right]\;. \
\label{vacmass4}
\eqa
We get $(m^m_\sigma)^2$,$(m^m_\eta)^2$,$(m^m_{a_0})^2$ and $(m^m_\eta)^2$ after substitution of the Eqs.(\ref{vacmass1})--(\ref{vacmass4}) into the Eq.(\ref{m2meson}) as
\bqa
\label{mesonm1}
(m^m_\sigma)^2&=&m^2_\sigma-\dfrac{N_cg^4\overline{\sigma}^2}{2(4\pi)^2}\left[1+3\ln\left(\dfrac{\Lambda^2}{m^2_q}\right)\right]\;,
\eqa
\bqa
(m^m_{a_0})^2&=&m^2_{a_0}-\dfrac{N_cg^4\overline{\sigma}^2}{2(4\pi)^2}\left[1+3\ln\left(\dfrac{\Lambda^2}{m^2_q}\right)\right]\;,
\eqa
\bqa
(m^m_\eta)^2&=&m^2_\eta-\dfrac{N_cg^4\overline{\eta}^2}{2(4\pi)^2}\left[1+\ln\left(\dfrac{\Lambda^2}{m^2_q}\right)\right]\;,
\eqa
\bqa
(m^m_\pi)^2&=&m^2_\pi-\dfrac{N_cg^4\overline{\pi}^2}{2(4\pi)^2}\left[1+\ln\left(\dfrac{\Lambda^2}{m^2_q}\right)\right]\;.
\label{mesonm4}
\eqa

The parameters in vacuum are obtained as

\bqa
\label{paraqmvt1}
\lambda_1&=&(m^m_\sigma)^2+(m^m_\eta)^2-(m^m_{a_0})^2-(m^m_\pi)^2\over 2 f^2_\pi
\eqa
\bqa
\lambda_2&=&(m^m_{a_0})^2-(m^m_\eta)^2\over f^2_\pi
\eqa
\bqa
m^2&=&(m^m_\pi)^2+\frac{(m^m_\sigma)^2-(m^m_\eta)^2}{2}
\eqa
\bqa
c&=&\frac{(m^m_\eta)^2-(m^m_\pi)^2}{2}
\label{paraqmvt4}
\eqa
we get the parameters of the QMVT on substitution of the Eqs.(\ref{mesonm1})--(\ref{mesonm4}) into the Eqs.(\ref{paraqmvt1})--(\ref{paraqmvt4}) and found that $\lambda_1$,$c$ of the QMVT are same with respect  to $\lambda_1$,$c$ of the QM.We observe change in $\lambda_2$ and $m^2$ as,
\bqa
\lambda_2&=&\lambda_{2s}-\dfrac{N_cg^4}{(4\pi)^2}\ln\left(\dfrac{4\Lambda^2}{g^2f^2_\pi}\right)\;
\eqa
\bqa
m^2&=&m^2_{s}-\dfrac{N_cg^4f^2_\pi}{2(4\pi)^2}\;.
\eqa
where $\lambda_{2s}$ and $m^2_s$ are same old $\lambda_{2}$ and $m^2$ parameters of the QM model.

\section{INTEGRALS AND SUM INTEGRALS}
\label{appenB}
The divergent loop integrals are regularized by encorporating dimensional regularization.
\bqa
\int_p=\left(\frac{e^{\gamma_E}\Lambda^2}{4\pi}\right)^\epsilon\int \frac{d^dp}{(2\pi)^d}\;,
\eqa
where $d=4-2\epsilon$ , $\gamma_E$ is the Euler-Mascheroni constant, and $\Lambda$ is renormalization scale associated with the $\overline{\text{MS}}$.

\bqa
\nonumber
\mathcal{A}(m^2_q)&=&\int_p \frac{1}{p^2-m^2_q}=\frac{i m^2_q}{(4\pi)^2}\left[\frac{1}{\epsilon}+1\right. \\
\nonumber
&&\left.+\ln(4\pi e^{-\gamma_E})+\ln\left(\frac{\Lambda^2}{m^2_q}\right)\right]\;
\eqa

we rewrite this after redefining $\Lambda^2\longrightarrow \Lambda^2\frac{e^{\gamma_E}}{4\pi}$.

\bqa
\label{aint1}
\mathcal{A}(m^2_q)&=&\frac{i m^2_q}{(4\pi)^2}\left[\frac{1}{\epsilon}+1+\ln\left(\frac{\Lambda^2}{m^2_q}\right)\right]
\eqa

\bqa
\label{bint1}
\nonumber
\mathcal{B}(p^2)&=&\int_k \frac{1}{(k^2-m^2_q)[(k+p)^2-m^2_q)]} \\
&=&\frac{i}{(4\pi)^2}\left[\frac{1}{\epsilon}+\ln\left(\frac{\Lambda^2}{m^2_q}\right)+C(p^2)\right]
\eqa

\bqa
\label{bprimeint1}
\mathcal{B}^\prime(p^2)&=&\frac{i}{(4\pi)^2}C^\prime(p^2)
\eqa

\begin{widetext}
\begin{eqnarray}
\label{eq:cp}
C(p^2)&=& 
\begin{dcases}
2-2\sqrt{\dfrac{4 m^2_q}{p^2}-1}\arctan\left(\dfrac{1}{\sqrt{\dfrac{4 m^2_q}{p^2}-1}}\right)\;,
& (p^2 < 4 m^2_q)\\ 
2+\sqrt{1-\dfrac{4 m^2_q}{p^2}}\ln\left(\dfrac{1-\sqrt{1-\dfrac{4 m^2_q}{p^2}}}{1+\sqrt{1-\dfrac{4 m^2_q}{p^2}}}\right), & (p^2 > 4 m^2_q) 
\end{dcases}
\end{eqnarray}

\begin{eqnarray}
\label{eq:cpp}
C^{\prime}(p^2)&=& 
\begin{dcases}
\frac{4 m^2_q}{p^4\sqrt{\dfrac{4 m^2_q}{p^2}-1}}\arctan\left(\dfrac{1}{\sqrt{\dfrac{4 m^2_q}{p^2}-1}}\right)-\frac{1}{p^2}\;,                       & (p^2 < 4 m^2_q)\\ 
\frac{2 m^2_q}{p^4\sqrt{\dfrac{4 m^2_q}{p^2}-1}}\ln\left(\dfrac{1-\sqrt{1-\dfrac{4 m^2_q}{p^2}}}{1+\sqrt{1-\dfrac{4 m^2_q}{p^2}}}\right)-\frac{1}{p^2}, & (p^2 > 4 m^2_q) 
\end{dcases}
\end{eqnarray}
\end{widetext}


\bibliography{refs}{}

\bibliographystyle{apsrmp4-1}

\end{document}